\documentclass[twocolumn]{aastex631}
\pdfoutput=1

\let\tablenum\relax
\usepackage[separate-uncertainty=true, range-phrase=--, range-units=single,
angle-symbol-over-decimal, retain-explicit-plus]{siunitx}
\usepackage{amsmath}
\usepackage{graphicx}
\usepackage{chemformula}
\usepackage{acro}
\usepackage{multirow}
\usepackage[normalem]{ulem}

\received{}

\shorttitle{}
\shortauthors{}

\DeclareSIUnit{\mag}{mag}
\DeclareSIUnit{\pixel}{pixel}
\DeclareSIUnit{\parsec}{pc}
\DeclareSIUnit{\arcsec}{arcsec}
\DeclareSIUnit{\arcmin}{arcmin}
\DeclareSIUnit{\solarlum}{\mbox{$L_\odot$}}
\DeclareSIUnit{\solarmass}{\mbox{$M_\odot$}}
\DeclareSIUnit{\radius}{\mbox{$R_{25}$}}
\DeclareSIUnit{\perradius}{\mbox{$R^{-1}_{25}$}}
\DeclareSIUnit{\year}{yr}
\DeclareSIUnit{\deg}{deg}

\DeclareSIUnit{\adu}{ADU}
\DeclareSIUnit{\erg}{erg}
\DeclareSIUnit{\dex}{dex}

\newcommand{\ha}{H$\alpha$}
\newcommand{\hb}{H$\beta$}
\newcommand{\oiii}{[\ion{O}{3}]}
\newcommand{\nii}{[\ion{N}{2}]}
\newcommand{\oii}{[\ion{O}{2}]}
\newcommand{\sii}{[\ion{S}{2}]}
\newcommand{\hii}{\ion{H}{2}}

\DeclareAcroEnding{possessive}{'s}{'s}

\NewAcroCommand\acg{m}{\acropossessive\UseAcroTemplate{first}{#1}}

\DeclareAcronym{EW}{short=EW, long=equivalent width}
\DeclareAcronym{2MASS}{short=2MASS, long=Two Micron All Sky Survey}
\DeclareAcronym{cigale}{short=\textsc{cigale}, long=Code Investigating GALaxy Emission}
\DeclareAcronym{chaos}{short=CHAOS, long=CHemical Abundances of Spirals}
\DeclareAcronym{SFR}{short=SFR, long=star formation rate}
\DeclareAcronym{psf}{short=PSF, long=point spread function}
\DeclareAcronym{agn}{short=AGN, long=active galactic nucleus, long-plural-form=active galactic nuclei}
\DeclareAcronym{XUV}{short=XUV, long=extended-UV}
\DeclareAcronym{SFE}{short=SFE, long=star formation efficiency}
\DeclareAcronym{LSB}{short=LSB, long=low surface brightness}
\DeclareAcronym{ISM}{short=ISM, long=interstellar medium}
\DeclareAcronym{ACS}{short=ACS, long=Advanced Camera for Surveys}
\DeclareAcronym{WFC3}{short=WFC3, long=Wide Field Camera 3}
\DeclareAcronym{sdss}{short=SDSS, long=Sloan Digital Sky Survey}
\DeclareAcronym{cwru}{short=CWRU, long=Case Western Reserve University}
\DeclareAcronym{aic}{short=AIC, long=Akaike Information Criterion}
\DeclareAcronym{bic}{short=BIC, long=Bayesian Information Criterion}

\defcitealias{croxall2016}{CHAOS}

\begin{document}

\title{Deep Narrowband Photometry of the M101 Group: Strong-Line Abundances of 720 \hii\ Regions}
\author{Ray Garner, III}
\affiliation{Department of Astronomy, Case Western Reserve University, 10900 Euclid Avenue, Cleveland, OH 44106, USA}

\author{J. Christopher Mihos}
\affiliation{Department of Astronomy, Case Western Reserve University, 10900 Euclid Avenue, Cleveland, OH 44106, USA}

\author{Paul Harding}
\affiliation{Department of Astronomy, Case Western Reserve University, 10900 Euclid Avenue, Cleveland, OH 44106, USA}

\author{Aaron E. Watkins}
\affiliation{Centre for Astrophysics Research, School of Physics, Astronomy \& Mathematics, University of Hertfordshire, Hatfield, AL10 9A, UK}

\author{Stacy S. McGaugh}
\affiliation{Department of Astronomy, Case Western Reserve University, 10900 Euclid Avenue, Cleveland, OH 44106, USA}

\begin{abstract}

We present deep, narrowband imaging of the nearby spiral galaxy M101 and its satellites to analyze the oxygen abundances of their \hii\ regions. Using CWRU's Burrell Schmidt telescope, we add to the narrowband dataset of the M101 Group, consisting of \ha, \hb, and \oiii\ emission lines, the blue \oii$\lambda3727$ emission line for the first time. This allows for complete spatial coverage of the oxygen abundance of the entire M101 Group. We used the strong-line ratio $R_{23}$ to estimate oxygen abundances for the \hii\ regions in our sample, utilizing three different calibration techniques to provide a baseline estimate of the oxygen abundances. This results in $\sim$650  \hii\ regions for M101, 10 \hii\ regions for NGC~5477, and $\sim$60 \hii\ regions for NGC~5474, the largest sample for this Group to date. M101 shows a strong abundance gradient while the satellite galaxies present little or no gradient. There is some evidence for a flattening of the gradient in M101 beyond $R \sim \SI{14}{\kilo\parsec}$.  Additionally, M101 shows signs of azimuthal abundance variations to the west and southwest. The radial and azimuthal abundance variations in M101 are likely explained by an interaction it had with its most massive satellite NGC~5474 $\sim$\SI{300}{\mega\year} ago combined with internal dynamical effects such as corotation. 

\end{abstract}

\section{Introduction}\label{sec:intro}

Star forming regions provide a tracer of galactic chemical evolution through the gas-phase metallicity, typically measured in \hii\ regions by the ratio of oxygen to hydrogen atoms. This oxygen is synthesized in high-mass stars and released to the \ac{ISM} through stellar winds or supernovae. It is then mixed throughout a galaxy through a variety of mixing processes, such as mergers, turbulence, and instabilities in the disk (see \citealt{roy1995}). 

In the past, most studies of the oxygen abundances\footnote{In this paper ``metallicity'' and ``oxygen abundance'' will be used interchangeably unless otherwise noted.} of \hii\ regions have revealed the presence of monotonically decreasing radial gradients, typically around $-\SI{0.05}{\dex\per\kilo\parsec}$ \citep[e.g.][]{vilacostas1992,zaritsky1994,vanzee1998}, i.e., the inner regions of galaxies are more metal-rich than the outskirts. These gradients were easily explained as a consequence of the inside-out galaxy formation scenario \citep[e.g.][]{scannapieco2009}. However, the presence of breaks in the oxygen abundance gradients have been observed in numerous galaxies, including M101 \citep{vilacostas1992,zaritsky1992}. 

Traditionally, oxygen abundance breaks, beyond which the abundance gradient of a galaxy flattens, have only been detected by using the so-called ``strong-line'' methods, those methods that relate the oxygen abundance to the strong recombination and collisionally-excited (forbidden) lines (e.g., \oii$\lambda3727$, \hb, \oiii$\lambda\lambda4959,5007$, \ha). These methods are used when auroral lines, like \oiii$\lambda4363$, are undetectable and don't allow for the electron temperature of a gas to be estimated, called the ``direct'' method \citep{dinerstein1990,skillman1998,stasinska2007}. Although the strong-line methods have gained widespread use with their easily measurable lines, they come with their own host of issues \citep[see, e.g.,][for a discussion]{pilyugin2003,kewley2008}. It was these defects that were blamed for producing ``spurious'' breaks where none existed. 

However, with the creation of newer strong-line abundance methods, breaks in oxygen abundance gradients have been reintroduced to the literature. Initially, only galaxies with \ac{XUV} emission \citep{thilker2007} were found to have flattened gradients, such as M83 \citep{bresolin2009} or NGC~4625 \citep{goddard2011}. While these results were produced using strong-line abundances, the introduction of integral field spectroscopy allowed direct methods to be applied across entire galaxy disks showing that many non-\ac{XUV} galaxies have flattened radial abundances beyond their $R_{25}$ \citep{sanchez2014,sanchezmenguiano2016}. Even our own Galaxy has some evidence of abundance flattening, measured not from \hii\ regions but from the [Fe/H] ratio of open clusters \citep{lepine2011} and the [O/H] ratio of OB stars \citep{daflon2004}. 

Numerous physical mechanisms have been proposed to explain the flattening of abundance gradients at large radii. Some attribute it to dynamical effects, such as the barrier effect of corotation, which isolates the inner and outer regions of a galaxy disk from each other \citep{lepine2011,scarano2013}, or other resonances with spiral density waves \citep{sellwood2002}. Others attribute it to an environmental cause such as minor mergers or satellite interactions increasing the gas content in the outer disk \citep{bird2012,lopezsanchez2015}. There is also the possibility of a different star formation efficiency in the outer disk \citep{bigiel2010,bresolin2012} as would be predicted in a nonlinear Kennicutt-Schmidt law \citep{esteban2013}. Whatever the cause, it is clear that the chemical evolution of a galaxy is tied to its dynamical evolution. 

In addition to breaks in radial abundance gradients, there are also some suggestions that there should be azimuthal variations across galaxy disks as well. This is predicted to be caused by the different scales on which chemical enrichment operates \citep{roy1995}. Given the large difference in timescales necessary to produce oxygen (OB star lifetime; $<$\SI{10}{\mega\year}) and mixing on sub-kiloparsec scales (\SIrange{10}{100}{\mega\year}; \citealt{roy1995}), there should be some azimuthal inhomogeneity of oxygen abundance. Tentative evidence has been found in the Milky Way via the [Fe/H] ratio in Cepheids \citep{pedicelli2009,lepine2011,genovali2014}, as well as in M101 via oxygen abundances \citep{kennicutt1996,li2013}. Despite this, detecting azimuthal variations in other galaxies has been difficult because spectroscopic studies are usually limited to bright \hii\ regions resulting in biased coverage. 

In an effort to add to the growing evidence of radial breaks and azimuthal variations in the oxygen abundances of galaxies, we have used \acg{cwru} Burrell Schmidt 24/36-inch telescope to add to the deep, wide-field, multiline, narrowband observations of the nearby spiral galaxy M101 (NGC~5457) and its group environment presented in previous papers \citep{watkins2017,garner2021}. In addition to the previous narrowband images targeting \ha, \hb, and \oiii\ \citep{watkins2017,garner2021}, we now add deep narrowband images targeting \oii\ for the first time. This allows us to sample the entire disks of M101 and its major satellites, NGC~5477 and NGC~5474 (see Table~\ref{props}), with our large survey area ($\sim$\SI{6}{\square\deg}) and our photometric depth which allows us to analyze faint inter-arm and outer disk \hii\ regions. 

M101 was chosen for this survey because its nearby distance ($D = \SI{6.9}{\mega\parsec}$; see \citealt{matheson2012} and references therein) enables its properties to be studied in detail. Notably, given its close distance and subsequently high spatial resolution, even the faintest low-luminosity \hii\ regions can be resolved.   M101 is also currently interacting with its satellite population, likely the massive satellite NGC~5474 \citep{beale1969,rownd1994,waller1997,linden2022}. Given that interacting systems have systematically lower oxygen abundances than isolated galaxies and sometimes have flattened gradients \citep{kewley2006}, and that M101 has been heavily studied by spectroscopic surveys \citep[e.g.][]{kennicutt1996,kennicutt2003,li2013,croxall2016,hu2018}, the M101 Group is an excellent group for studying the chemical evolution of interacting galaxies.

\begin{deluxetable}{lccc}
\tablecaption{Properties of M101 and Nearby Companions \label{props}} 
\tablehead{ & \colhead{M101} & \colhead{NGC~5477} & \colhead{NGC~5474}}
\startdata
R.A. & 14:03:12.5 & 14:05:33.3 & 14:05:01.6 \\
Decl. & +54:20:56 & +54:27:40 & +53:39:44 \\
Type & S\underline{A}B(rs)c & dIm & SA(s)m \\
$V_{\text{helio}}$ [\si{\kilo\metre\per\second}] & 241 & 304 & 273 \\
Distance [\si{\mega\parsec}] & 6.9 & \dots & \dots \\
$R_{\text{proj}}$ [\si{\arcmin}] & \dots & 22 & 44 \\
$R_{\text{proj}}$ [\si{\kilo\parsec}] & \dots & 44 & 88 \\
$R_{25}$ [\si{\arcmin}] & 8.0 & 0.8 & 2.4 \\
$R_{25}$ [\si{\kilo\parsec}] & 16.0 & 1.6 & 4.8
\enddata
\tablecomments{Hubble types are taken from the CVRHS system of \citet{buta2015} and the sizes of NGC~5477 and NGC~5474 are taken from the RC3 \citep{devaucouleurs1991}, while the size of M101 is taken from \citet{mihos2013}. The physical distance to M101 is adopted from \citet{matheson2012}, and references therein. $R_{\text{proj}}$ is the projected distance from M101.}
\end{deluxetable}

\section{Observations and Data Reduction}\label{sec:obs}

The narrowband imaging data for this project was taken using \ac{cwru}
Astronomy's 24/36-inch Burrell Schmidt telescope located at Kitt Peak
Observatory. Full details of our narrowband imaging techniques are given
in \citet{watkins2017}, and summarized briefly here.
Quantitative information about the narrowband filters and final image stacks for
our imaging dataset is given in Table~\ref{narrowbandobs}.

Over the course of four observing seasons, we observed M101 in a set of
narrow on-band filters ($\Delta\lambda\approx$\SIrange{80}{100}{\angstrom}) centered on
the red-shifted emission lines \ha$\lambda$6563, \hb$\lambda$4861, \oiii$\lambda\lambda$4959,5007, and
\oii$\lambda\lambda$3727,3729. To measure adjacent stellar continuum
for each line, we also observed the galaxy in narrow off-band filters
shifted in wavelength by $\approx$\SIrange{100}{150}{\angstrom} from each on-band
filter. The Burrell Schmidt images a $\ang{1.65}\times\ang{1.65}$ field of
view onto a $4096\times4096$ back-illuminated CCD, yielding a pixel scale of
\SI{1.45}{\arcsec\per\pixel}. In each filter, we took \numrange{50}{70} \SI{1200}{\second} images
of the galaxy, randomly dithering the telescope by \SIrange{10}{30}{\arcmin}
between exposures. We observed on photometric nights with no moon,
yielding night sky levels of \SIrange{50}{100}{\adu\per\pixel}. We built flat
fields in each filter from a combination of twilight flats and stacked
night sky exposures \citep[see][]{watkins2017}, and constructed a spatially dependent scattered
light model for the degree-scale \ac{psf} of bright stars using deep \SI{1200}{\second}
observations of the bright stars Regulus and Arcturus in each filter.
Finally, to assist in flux calibration, we also observed
spectrophotometric standards \citep{massey1988} throughout the course of each
observing run.

During data reduction, we first subtracted from each image a
median-combined nightly master bias frame, then flat-fielded the images
using the composite twilight/dark-sky flat field. We then removed
scattered light from bright stars in the field using our degree-scale
\ac{psf} model (following the technique of \citealt{slater2009}), after which we model
and subtract the sky background using a plane fit to blank sky pixels in
each image. We then register and median-stack the images to create final
on- and off-band composite images in each filter, with total exposure
times varying from \SIrange{18}{24}{\hour} each. 

We flux calibrate the final image stacks using a variety of techniques,
intercomparing the results to estimate our final flux zeropoint
uncertainties. We calibrate by (1) deriving a photometric
solution from observations of \citet{massey1988} spectrophotometric standard
stars, (2) measuring zeropoints from $ugr$ magnitudes of the $\sim$150 stars in the M101 field from the Sloan Digital Sky Survey (\acs{sdss}; and
including a synthesized broadband--narrowband color term for each
filter), and (3) synthesizing narrowband magnitudes using
\ac{sdss} spectroscopy of $\sim$100 point sources in the M101 field. These
different techniques yielded flux zeropoints that agreed to within $\pm$\SI{5}{\percent},
which we take to be our absolute flux uncertainties in the imaging data.

Table~\ref{narrowbandobs} also details the 1$\sigma$ limiting depth of the images,
measured in two ways. First, the per-pixel noise level ($\sigma_{\rm pix}$) in
each image is measured using the standard deviation of pixel intensities
measured within regions of blank sky around M101. Second, the limiting
depth over larger scales is measured from the scatter in mean flux
measured with $\ang{;;30}\times\ang{;;30}$ ($20\times20$ pixel) boxes
around M101 ($\sigma_{30}$), and traces our sensitivity to diffuse
extended narrowband emission.

\begin{deluxetable*}{ccccccccc}
\tablewidth{0pt}
\tablecaption{Narrowband Imaging Datasets}
\tablehead{\colhead{Season} &\colhead{Filter} & \colhead{$\lambda_0$} & \colhead{$\Delta\lambda$} &
\colhead{Exposure Time} & \colhead{ZP (flux)} & \colhead{ZP (AB)} &
\colhead{$\sigma_{\text{pix}}$} & \colhead{$\sigma_{30}$}
}
\startdata
2014 & \ha-on     & \SI{6590}{\angstrom} & \SI{101}{\angstrom} & $71 \times \SI{1200}{\second}$ & \num{5.61e-18} & 26.63 & \num{3.4e-18} & \num{1.2e-18} \\
2014 & \ha-off    & \SI{6726}{\angstrom} & \SI{104}{\angstrom} & $71 \times \SI{1200}{\second}$ & \num{5.50e-18} & 26.64 & \num{2.8e-18} & \num{1.3e-18} \\ 
2018 & \hb-on     & \SI{4875}{\angstrom} & \SI{82}{\angstrom} & $59 \times \SI{1200}{\second}$ & \num{7.65e-18} & 26.73 & \num{2.9e-18} & \num{7.3e-19} \\
2018 & \hb-off    & \SI{4757}{\angstrom} & \SI{81}{\angstrom} & $55 \times \SI{1200}{\second}$ & \num{7.91e-18} & 26.74 & \num{2.9e-18} & \num{5.7e-19} \\
2019 & \oiii-on  & \SI{5008}{\angstrom} & \SI{102}{\angstrom} & $67 \times \SI{1200}{\second}$ & \num{7.58e-18} & 26.91 & \num{3.0e-18} & \num{7.4e-19} \\
2019 & \oiii-off & \SI{5114}{\angstrom} & \SI{101}{\angstrom} & $66 \times \SI{1200}{\second}$ & \num{7.37e-18} & 26.89 & \num{3.0e-18} & \num{7.9e-19} \\
2021 & \oii-on   & \SI{3747}{\angstrom} & \SI{79}{\angstrom} & $60 \times \SI{1200}{\second}$ & \num{1.74e-17} & 26.37 & \num{5.3e-18} & \num{8.7e-19} \\
2021 & \oii-off  & \SI{3839}{\angstrom} & \SI{80}{\angstrom} & $51 \times \SI{1200}{\second}$ & \num{1.48e-17} & 26.50 & \num{4.8e-18} & \num{8.0e-19} \\
\enddata
\tablecomments{ZP (flux) converts \SI{1}{\adu} to \si{\erg\per\second\per\square\centi\metre} in the master
images, while ZP (AB) converts to AB magnitudes. $\sigma_{\text{pix}}$ is the per-pixel standard deviation
measured in \si{\erg\per\second\per\square\centi\metre\per\square\arcsec}, while $\sigma_{30}$ is the 1$\sigma$ variation
in mean intensity measured in \ang{;;30}$\times$\ang{;;30} blank sky boxes, in the same units.}
\label{narrowbandobs}
\end{deluxetable*}

We stress that as noted in Table~\ref{props}, throughout this paper we use a different value for M101's $R_{25} = \ang{;8;}$ \citep{mihos2013} than the usual RC3 value of $R_{25} = \ang{;14.4;}$ \citep{devaucouleurs1991}. As noted by \citet{mihos2013}, a small patch of starlight to the north of M101's main disk and at the end of the galaxy's northeastern arm is likely to blame for the large radius measured by \citet{devaucouleurs1991} despite most of the galaxy not extending that far. We use the areal-weighted $R_{25}$ measured by \citet{mihos2013} as a more robust estimate of the galaxy's size. When comparing our data to spectroscopic data, we will convert their scaled radii to our scale. This has the effect of roughly doubling the scaled radii and halving the abundance gradients reported in the original papers. 

\section{Selecting and Correcting \hii\ Regions}

\subsection{\hii\ Region Detection}

Our final imaging dataset from the Burrell Schmidt consists of the narrowband imaging described in Section \ref{sec:obs}, as well as deep broadband imaging in (modified) Johnson $B$ and Washington $M$ filters from \citet{mihos2013}. The broadband photometry has been transformed to standard Johnson $B$ and $V$ as described in \citet{mihos2013}. Our general technique for detecting \hii\ regions in M101 and its satellites is much the same as utilized in \cite{garner2021} with some key differences described below to account for region detection within a galaxy rather than interspersed throughout the M101 Group. 

However, as described in detail by \cite{garner2021} and noted elsewhere \citep[e.g.][]{kellar2012,watkins2017}, stars will broadly mimic \ha\ emission as our off-band filters can sample absorption features in the stellar continuum. In order to mask the contribution of these stars, we utilized two surveys targeting both bright and faint stars. We masked those stars in the Tycho-2 Catalog \citep{hog2000} brighter than $B_T = 12.5$. All of these stars were found outside the disks of the galaxies and do not affect our analysis. For fainter stars not in the Tycho-2 Catalog, we used entries from the \ac{2MASS} All-Sky Catalog of Point Sources \citep{skrutskie2006}, queried through the VizieR interface of \texttt{astropy}'s \texttt{Astroquery} package \citep{ginsburg2019}. Within a $\ang{;30;} \times \ang{;30;}$ cutout centered on M101, this resulted in \num{421} point sources being detected. In \citet{garner2021} we used \emph{Gaia} data to remove stars from our imaging of the intragroup field around M101, but here we are working within the bright disk of M101 where such data is not available, thus the necessity of using \ac{2MASS} photometry instead.

Given this list of point sources, we then make photometric cuts to determine whether a particular point source is a star or an \hii\ region. Ideally, we could use the \ac{2MASS} $JHK_s$ photometry provided in the Point Source Catalog to determine which sources are stars as they should follow a particular path through a $J-H$ vs.\ $H-K_s$ color-color plot \citep{straizys2009}. Since many of these point sources lie within or very close to M101, the \ac{2MASS} photometry has significant uncertainties due to the high and variable background. Since \ac{2MASS} does not provide a local background estimate, we instead perform aperture photometry on our $B$, $V$, and \ha\ on- and off-band images using \ang{;;8.7} apertures, or six times the image pixel scale. 

\begin{figure}
\plotone{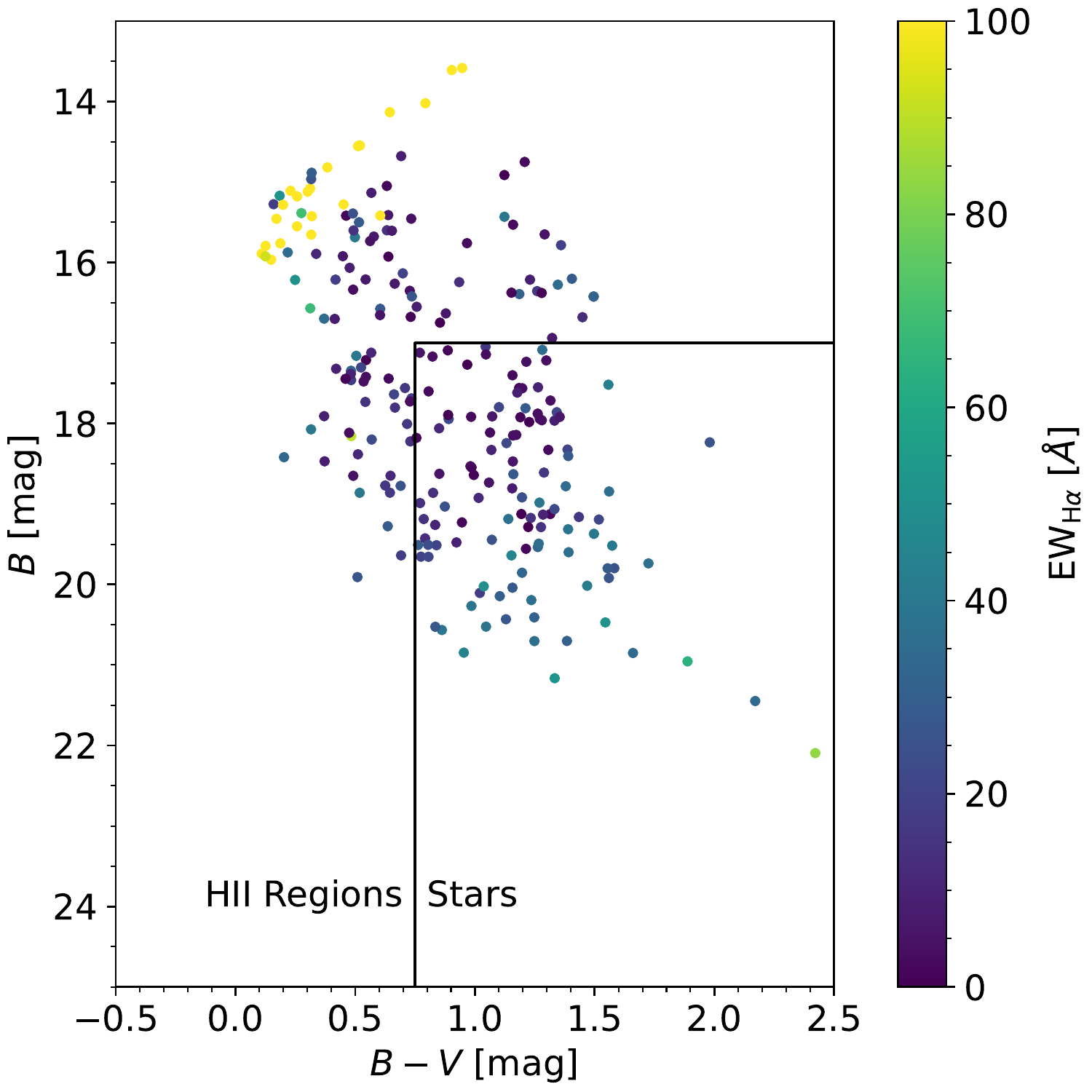}
\caption{A color-magnitude diagram for sources in the \acs{2MASS} All-Sky Catalog of Point Sources \citep{skrutskie2006} within a $\ang{;30;} \times \ang{;30;}$ cutout centered on M101. Sources are colored according to their \ha\ \acs*{EW}. The box indicated by the solid black lines indicates those sources that are consistent with stars that we subsequently mask in our segmentation routine. \label{star_selection}}
\end{figure}

Figure \ref{star_selection} shows the color-magnitude diagram for the \ac{2MASS} point sources with the marker color indicating their \ha\ \ac{EW} as measured from our narrowband imaging. Star-forming regions should be photometrically blue and have high \ha\ \ac{EW} (although older \hii\ regions or regions with a strong underlying stellar population can have slightly lower \acp{EW}), while stars should have progressively higher \ha\ \acp{EW} as a function of $B-V$ color, with red stars having the highest \ha\ \ac{EW} \citep[see Fig.\ 2 in][]{garner2021}. Using these photometric arguments, we define stars as having ($B-V > 0.75$ OR $B \geq 17$) AND \ha\ \ac{EW} $ < \SI{65}{\angstrom}$ as illustrated by the box outline in Figure \ref{star_selection}. 

Having masked both bright stars in the Tycho-2 Catalog around M101 and dim stars in the \ac{2MASS} Point Source Catalog in front of M101, we create a two-dimensional background object to calculate the background sky level and its uncertainty. As in \cite{garner2021}, all galaxies (both background and M101 Group members) were masked and the sky level was estimated in boxes of $100 \times 100$ pixels ($\ang{;;145} \times \ang{;;145}$) with filter sizes of $10 \times 10$ pixels ($\ang{;;14.5} \times \ang{;;14.5}$).

Then, we detected sources on the continuum-subtracted \ha\ image using \texttt{astropy}'s \texttt{PhotUtil} package and its \texttt{segmentation} module \citep{bradley2021}. This program detects sources as objects that have some minimum number of connected pixels, each greater than a background threshold value. In this case, the threshold value of $10\sigma$ above the background level on the continuum-subtracted \ha\ image described above after a two-pixel Gaussian smoothing was applied was chosen. This high threshold value was chosen in order to detect \emph{bona fide} star-forming regions within the galaxies in the M101 Group. As such, we do not mask the galaxies in our images when detecting these regions. 

This process resulted in \num{514} sources in M101 initially. We then deblended close or overlapping sources using 32 multi-thresholding levels and a contrast parameter of \num{0.015}, resulting in a final source list for M101 of \num{853} sources. We repeat this same process for both of the satellites in our sample, resulting in 6 and 51 sources before deblending and 11 and 71 sources after deblending for NGC~5477 and NGC~5474, respectively. 

\begin{figure*}
\plotone{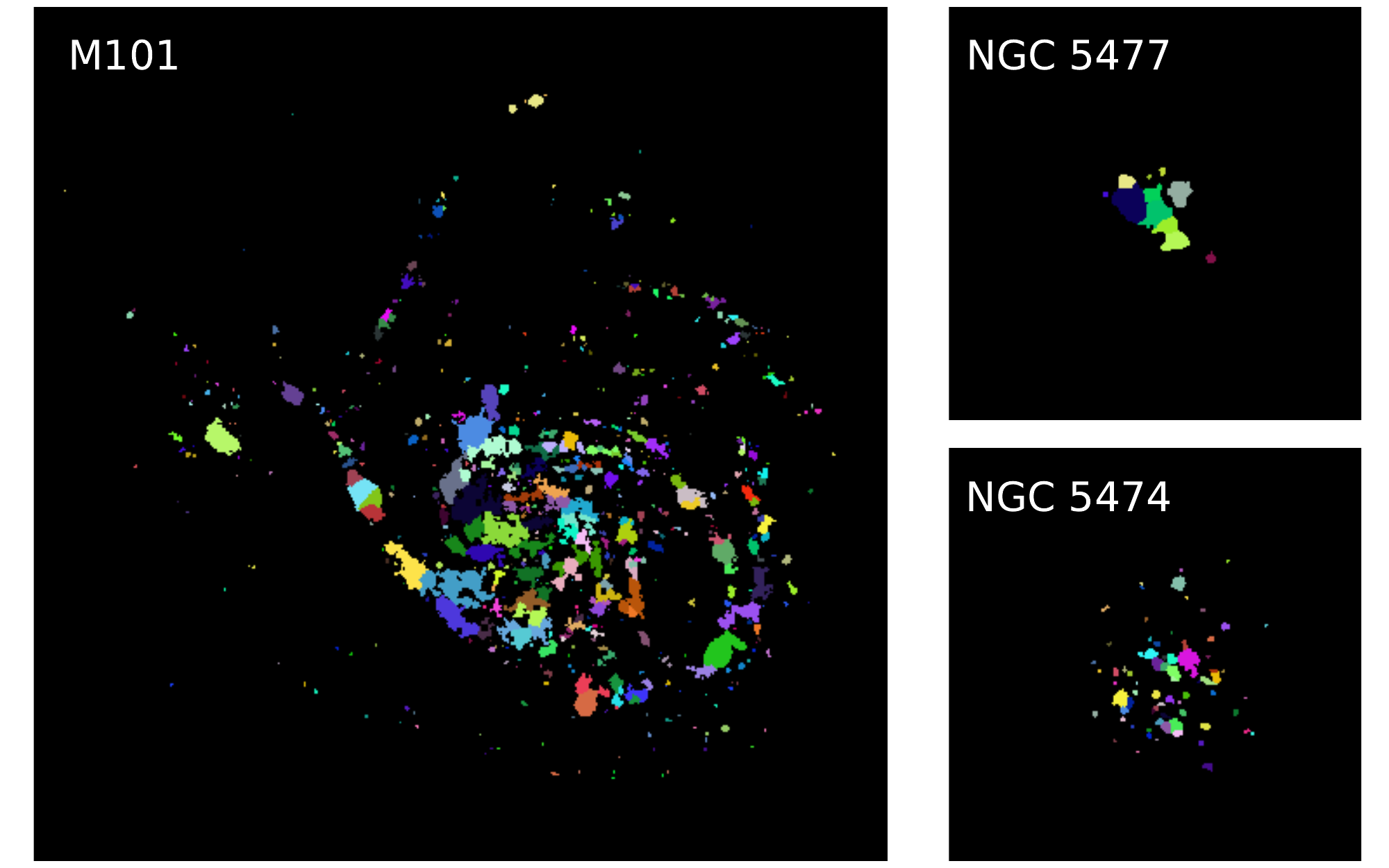}
\caption{The segmentation maps for M101 (left), NGC~5477 (top right), and NGC~5474 (bottom right). Each shaded region is a detected \hii\ region after star removal and deblending as described in the text. The M101 map measures $\ang{;30;} \times \ang{;30;}$, the NGC~5477 map measures $\ang{;5;} \times \ang{;5;}$, and the NGC~5474 map measures $\ang{;10;} \times \ang{;10;}$. In all images, north is up and east is to the left. \label{segm}}
\end{figure*}

With the segmentation maps shown in Figure \ref{segm} defining the \hii\ regions in each galaxy, we used the \texttt{PhotUtil} \texttt{segmentation} module to calculate photometric and structural quantities for each region in each of the narrowband images as well as in the broadband imaging of \cite{mihos2013}. Many of these were default calculations for \texttt{segmentation}, including positions and fluxes. We also calculated photometric errors, signal-to-noise, and AB magnitudes for each object in each filter. Additionally, we calculated emission line fluxes and \acp{EW} for each narrowband imaging pair (\ha, \hb, \oiii, and \oii), after correcting for a variety of photometric effects as detailed in the next section. 

\subsection{Photometric Corrections}\label{corrections}

\begin{figure}
\plotone{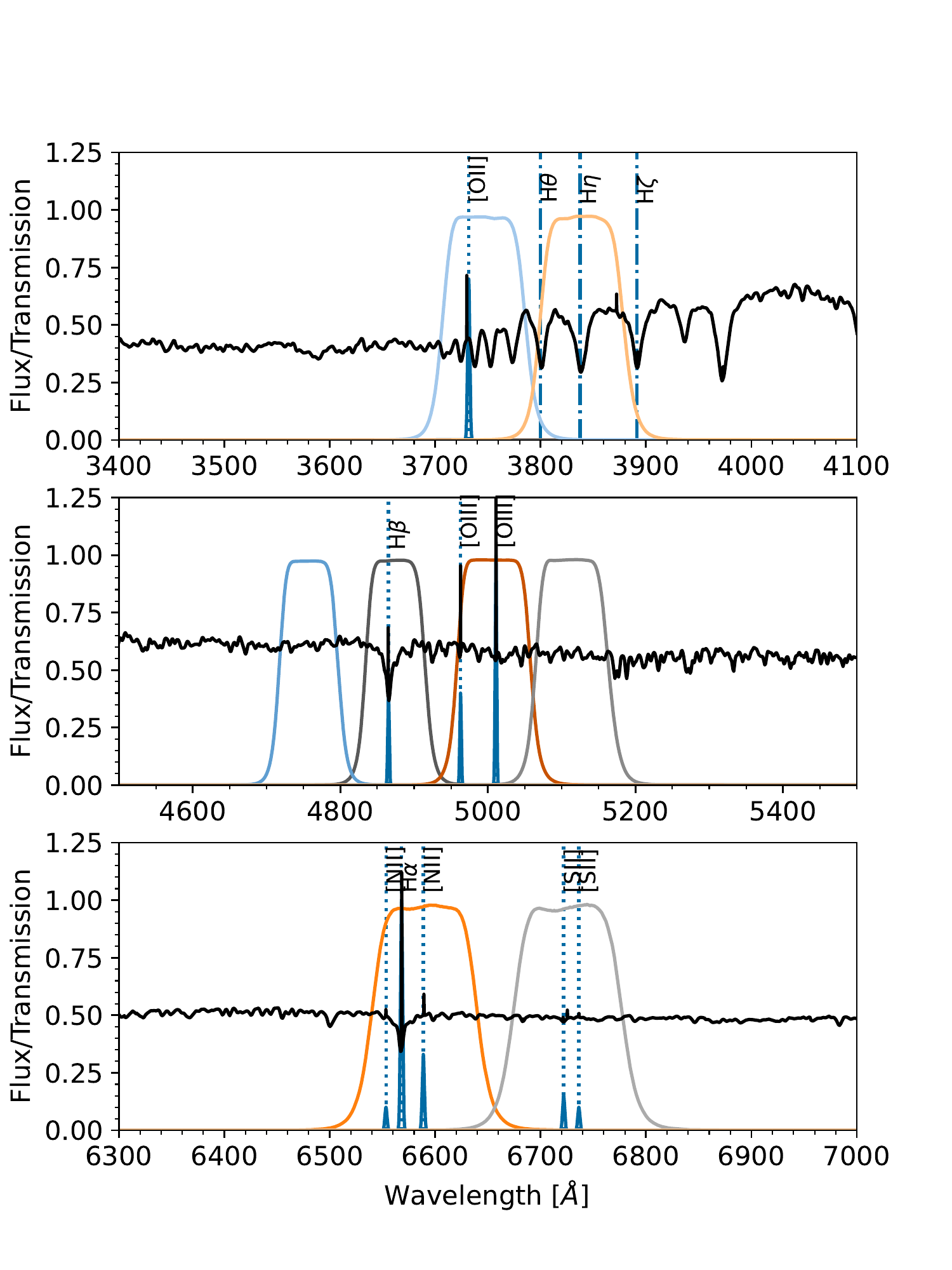}
\caption{The narrowband filters used to image the M101 Group overlaid on top of a synthetic \hii\ region spectrum from \acs{cigale}. Top: the \oii\ filters. Middle: the \hb\ and \oiii\ filters. Bottom: the \ha\ filters. In all cases, the on-band filter sits on top of their respective emission lines, while the off-band filter sits redward of the emission line except for the \hb-off filter which sits blueward. Particular emission and absorption lines are marked by vertical lines. \label{filters}}
\end{figure}

In narrowband photometry, since the off-band filter measures the strength of the continuum near each emission line captured in the on-band filters, theoretically, the flux differences calculated above are equivalent to each emission line's flux. However, there are several physical processes hindering that simple assumption. Figure \ref{filters} shows a synthetic, representative spectrum of a star-forming region generated with the Python Code Investigating GALaxy Emission (\acs{cigale}; \citealt{noll2009,boquien2019}) with particular emission and absorption lines marked. We plot our narrowband filter curves on top of the spectrum, showing that three of the four sets of narrowband filters have different problems that we need to correct: the \hb\ emission line strength is diminished by the effects of Balmer absorption in the stellar continuum; the \oii\ off-band filter does not sample a smooth continuum, instead sitting on high-order Balmer absorption; and the \ha\ on-band filter captures \ha+\nii\ emission while the \ha\ off-band filter captures \sii\ emission. Additionally, but not evident in Figure \ref{filters}, we must make a correction for any internal reddening. In this section, we describe the steps we take to correct for these various processes. 

In each \hii\ region, we correct the observed \hb\ \acp{EW} for underlying stellar absorption using the following technique. Each \hii\ region contains light from the young star-forming population along with an older underlying background stellar population. For the young population, we assume a stellar \hb\ \ac{EW} width of \SI{5}{\angstrom}, based on models and observations of \hii\ regions \citep[e.g.][]{gonzalezdelgado1999,gavazzi2004,moustakas2006}. For the underlying old stellar population, we adopt the \hb\ \ac{EW} measured in an annulus around each region with a size five to eight times the \hii\ region size. We then calculate a weighted average of these two \ac{EW} values, weighting by the off-band (i.e., stellar) flux ratio in the \hii\ region and surrounding annulus. This process yields net stellar absorption \ac{EW} corrections which fall in the range predicted by stellar population synthesis codes \citep[$\sim$\SIrange{2}{12}{\angstrom};][]{gonzalezdelgado1999}. In a small number of cases, typically in low signal-to-noise regions, the resulting \ac{EW} correction fell well outside the accepted ranges, and in these regions we simply assign an \hb\ \ac{EW} correction typical of other well-measured regions at similar radii in the galaxy. This process results in an average \hb\ \ac{EW} correction of \SI{4.0}{\angstrom}, \SI{2.2}{\angstrom}, and \SI{4.1}{\angstrom} for M101, NGC~5477, and NGC~5474, respectively. 

%

Higher order Balmer lines are increasingly sensitive to stellar absorption which gives rise to the absorption features seen in the \oii\ off-band filter in the top panel of Figure \ref{filters}. Here, the \oii\ off-band filter sits on H$\eta$ absorption and adjacent to H$\zeta$ and H$\theta$ absorption. \cite{gonzalezdelgado1999} used synthetic spectra to model the behavior of the H$\eta$ absorption line as a function of age and star formation history. They found that the H$\eta$ \ac{EW} ranges from \SIrange{2}{10}{\angstrom} with an average of \SI{5}{\angstrom}. However, we cannot make an \ac{EW} correction as we did to the \hb\ \acsp{EW} due to the complicated spectrum: the \oii\ off-band filter is sampling multiple emission and absorption lines and there is a rapidly changing continuum level between the on- and off-band filters. 

We use \acs{cigale} to model a range of star-forming histories for a short burst of star formation as well as those for an older, pre-existing stellar population. For the young population, we find typical on/off flux ratios of \num{1.00 \pm 0.02}, while for the older, pre-existing populations we find a lower ratio of \num{0.88 \pm 0.02}. For each region, we make the net correction based on a weighted combination of these two ratios, weighting by the off-band flux ratio of each \hii\ region and its surrounding environment in a similar fashion to what was done with the \hb\ correction.

Finally, we must make a correction for the fact that our \ha\ on-band filter measures \ha+\nii\ emission while the \ha\ off-band filter measures \sii\ emission. The relative \nii/\ha\ and \sii/\ha\ line ratios vary with metallicity, electron temperature, ionization state, etc., and not necessarily in the same way \citep{burbidge1962,baldwin1981,osterbrock1989}. Our previous work \citep{garner2021} corrected the \ha\ emission using a flux ratio of \nii/$\mathrm{H}\alpha = 0.33$ \citep{kennicutt1983,jansen2000}. However, \cite{james2005} demonstrated that this ratio works best for \nii-selected \hii\ regions, and overestimates the flux ratios of other regions and galaxies as a whole. 

\begin{figure}
\epsscale{1.2}
\plotone{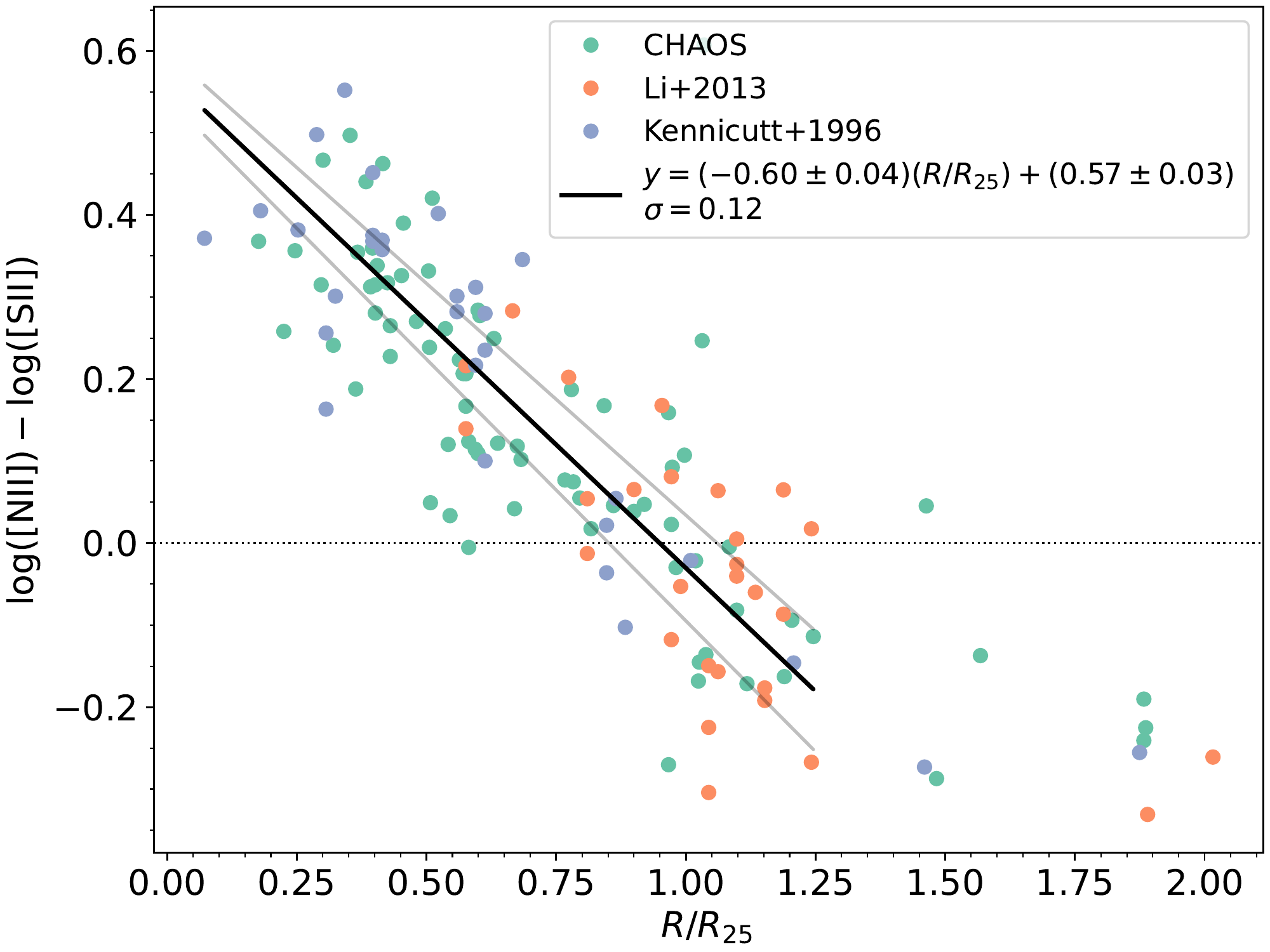}
\caption{The \nii/\sii\ residuals as a function of radius. Colored points indicate the survey: \citet[][green]{croxall2016}, \citet[][orange]{li2013}, and \citet[][blue]{kennicutt1996}. The solid black line indicates the best fit to the data within $1.25R_{25}$, with the adjoining solid gray lines showing the $1\sigma$ uncertainty. \label{line_comparison_radius}}
\end{figure}

Instead, we utilize published spectroscopic line ratios for \hii\ regions in M101 to determine the relation between \nii\ and \sii. In particular, we use the data of: \cite{kennicutt1996} totaling 41 regions; \cite{li2013} totaling 28 regions; and the \acl{chaos} group (\citealt{croxall2016}, hereafter CHAOS) totaling 77 regions. We investigated this data for a simple linear relation between \sii\ and \nii, but there was considerable scatter arguing that additional parameters may be in play. CHAOS found a strong radial gradient in the N/O ratio for M101 with a flattening at $R/R_{25} \gtrsim 1.35$, while there was no radial gradient in the S/O ratio. This suggests that there should be a radial gradient in the \nii/\sii\ ratio, which we show for the combined spectroscopic dataset in Figure \ref{line_comparison_radius}. Notably, this radial gradient flattens at $1.25R_{25}$, consistent with the radius noted by CHAOS above, where the ratio assumes a constant value of \nii/\sii $\approx10^{-0.25}$.

Therefore, we adopt the following \ha\ correction: 
\begin{equation}\label{ha_correction}
	f_{\mathrm{H}\alpha,\text{true}} = F_{\mathrm{H}\alpha,\text{diff}}\left(1 + \frac{f_{\text{[S \textsc{ii}]}}}{f_{\mathrm{H}\alpha}}\left(\frac{f_{\text{[N \textsc{ii}]}}}{f_{\text{[S \textsc{ii}]}}} - 1\right)\right)^{-1},
\end{equation}
where $f_{\mathrm{H}\alpha,\text{true}}$ is the true, corrected \ha\ flux, $F_{\mathrm{H}\alpha,\text{diff}}$ is the measured \ha\ flux difference, i.e.\ $F_{\mathrm{H}\alpha,\text{on}} - F_{\mathrm{H}\alpha,\text{off}}$, and $f_{\text{[S \textsc{ii}]}}/f_{\mathrm{H}\alpha} \approx \num{e-0.8}$ (CHAOS). In the above equation, we use the radial gradients in Figure \ref{line_comparison_radius} to determine the \nii/\sii\ ratio, where
\begin{equation}\label{ha_piecewise}
	\frac{f_{\text{[N \textsc{ii}]}}}{f_{\text{[S \textsc{ii}]}}} = \begin{cases}
		10^{-\num{0.60}(R/R_{25}) + \num{0.57}} & \text{if } R/R_{25} < 1.25 \\
		10^{-0.25} & \text{if } R/R_{25} \geq 1.25.
	\end{cases}
\end{equation}
This correction factor has the effect of reducing the observed \ha\ flux by $\sim$\SIrange{7}{25}{\percent} in M101. There is not enough data for the two satellite galaxies to perform a similar analysis. Instead, we employ a single correction to the data for each satellite. For NGC~5477, we calculate the correction based on the measured line strengths by \citet{berg2012}, while for NGC~5474, where no spectroscopic data is available, we use the mean correction for M101. These corrections amount to only a few percent in each case.

The last correction we make is the empirical extinction correction following the relation found in \citet{calzetti1994} and assuming the reddening curve found in \citet{calzetti2000}. This correction uses the Balmer decrement, \ha/\hb, to derive the extinction at \hb\ assuming Case B recombination ($T = \SI{e4}{\kelvin}, n_e = \SI{e2}{\per\cubic\centi\metre}$) so the intrinsic Balmer decrement is \ha/\hb $= 2.86$ \citep{osterbrock1989}. Given the measured extinction, applying this correction is straightforward and applied to the line fluxes and magnitudes in each band in each galaxy.

\begin{figure*}
\epsscale{1.2}
\plotone{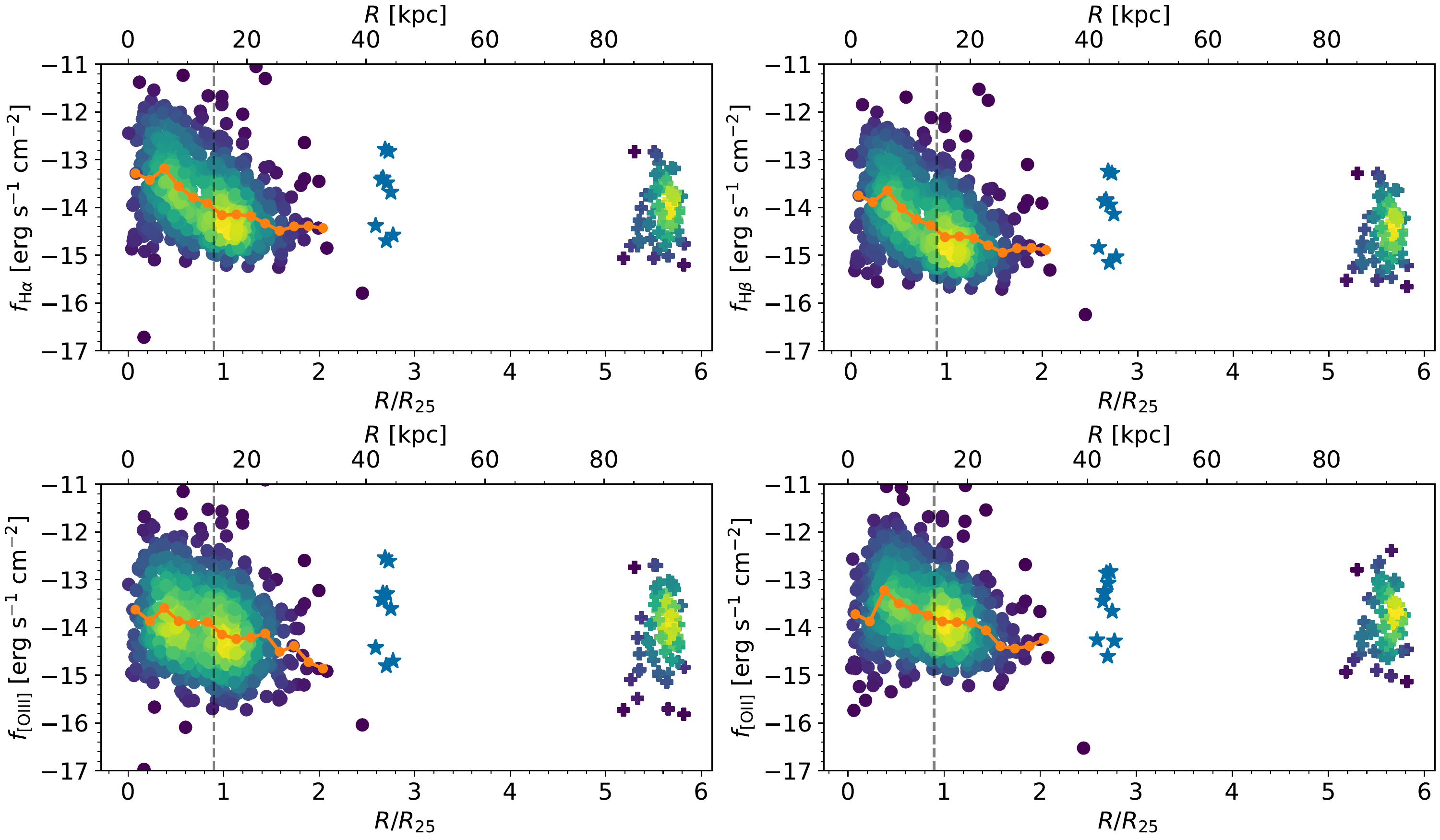}
\caption{The \ha, \hb, \oiii, and \oii\ fluxes plotted against radius from the center of M101. \hii\ regions in M101 are marked by circles and \hii\ regions in NGC~5474 are marked by plus signs; both are colored by local density of points in the plot. \hii\ regions in NGC~5477 are marked by blue stars. The orange lines are the median values for M101 in bins of $0.1R_{25}$, the standard deviation on each of the median values is $\sim$0.6, and the gray dashed lines mark the inner-outer disk of M101. In all plots, the corrections described in Section \ref{corrections} have been applied. \label{fluxes}}
\end{figure*}

Figure \ref{fluxes} shows radial profiles for each of the four (dereddened) line fluxes (\ha, \hb, \oiii, \oii) in each of the galaxies in our sample. \hii\ regions in M101 (circles) and NGC~5474 (plus signs) are colored by density of points, while those in NGC~5477 are simply marked by blue stars. The orange lines shows the median values for M101 in bins of $0.1R_{25}$ out to a radius of $2R_{25}$ and the dashed gray lines mark M101's outer disk, which we define at $>$3 times the azimuthally averaged disk scale length (\ang{;;430}, \SI{14.5}{\kilo\parsec}; \citealt{mihos2013,watkins2017}). Largely, we recover the trends shown in \cite{watkins2017}. Both NGC~5477's and NGC~5474's \hii\ regions span the same range of flux as those in M101's outer disk.  Additionally, the weak radial gradients in median \ha\ and \hb\ fluxes of the inner disk, and, to a lesser extent, \oiii\ and \oii\ fluxes, is most likely a demonstration of the so-called Kennicutt-Schmidt law: molecular gas density in M101 declines exponentially with radius \citep[e.g.][]{kenney1991}. However, the slight flattening we see in the outer disk could be evidence of different star formation efficiency; indeed, \citet{bigiel2010} showed that the star formation efficiency is considerably flatter in outer disks, between one and two isophotal radii.

Before estimating abundances and presenting the abundance gradients of the three galaxies in our sample, we briefly provide an overview of the numbers and photometric depth of our sample. In M101 (746 regions), we have reached a minimum photometric \ha\ flux of \SI{1.6e-16}{\erg\per\second\per\square\centi\metre}, which corresponds to an \ha\ luminosity of \SI{9.1e35}{\erg\per\second}. Using the SFR$-L(\text{\ha})$ calibration of \citet{kennicutt2012}, this corresponds to an \ac{SFR} of \SI{4.9e-6}{\solarmass\per\year}. In NGC~5477 (10 regions) and NGC~5474 (65 regions), the minimum photometric \ha\ flux is $\sim$\SI{e-15}{\erg\per\second\per\square\centi\metre}, which corresponds to an \ha\ luminosity of $\sim$\SI{e37}{\erg\per\second} and an \ac{SFR} of $\sim$\SI{e-5}{\solarmass\per\year}. We also quantify our lower photometric limits by calculating the total number of ionizing photons, $Q_0$, from \citet[][see Eqn.\ 1 in \citealt{garner2021}]{osterbrock1989}. While the faintest \hii\ regions in the satellite galaxies can be powered by approximately a single O7V star, the faintest region in M101 can be powered by approximately a single O9V star \citep{martins2005}. Compare these values to those obtained by CHAOS: 77 \hii\ regions targeted by spectroscopy, with a minimum \ha\ luminosity of \SI{2.3e36}{\erg\per\second}, an \ac{SFR} of \SI{1.2e-5}{\solarmass\per\year}, potentially powered by a single O8V star. 

\section{Cross-Comparison and Oxygen Abundances}

Having now corrected our data for the effects of underlying stellar absorption and contaminating emission lines, in this section we investigate the emission-line properties of our sample and calculate strong-line abundances. In Section \ref{sub:chaos}, we compare our narrowband photometry to that of high-quality spectroscopy performed by the CHAOS group. Finding our data matching the spectroscopic data within reasonable limits, in Section \ref{sub:ratios}, we investigate the radial trends between three strong-line ratios: the $R_{23}$ parameter \citep{pagel1979}, the ionization state given by O$_{32}$ \citep[e.g.][]{mcgaugh1991}, and the excitation state given by $P$ \citep{pilyugin2000,pilyugin2001}. Finally, in Section \ref{sub:oxy}, we derive oxygen abundances from our narrowband photometry using three different methods. 

\subsection{CHAOS Comparison}\label{sub:chaos}

To gauge the accuracy of our data reduction techniques described above in Section \ref{corrections}, we compare our photometric data with the spectroscopic data of the CHAOS group \citep{croxall2016}. We perform aperture photometry on the regions targeted by CHAOS in all of our narrowband images. Although the spectroscopic data was collected using slits with \ang{;;1} width, we use \ang{;;3} photometric apertures in our data, given its lower image resolution (\ang{;;2.2} FWHM and \ang{;;1.45}\,\si{\per\pixel}). This leads to a systematic bias in the comparison: we expect our fluxes to be brighter with respect to the CHAOS data, and the line ratios may show systematic differences as well due to gradients between neighboring individual \hii\ regions. 

With this noted, we perform standard aperture photometry on each of the 77 regions selected by CHAOS. For the \oii, \oiii, and \hb\ flux comparison, we calculate the line fluxes as outlined in Section~\ref{corrections}. However, for \ha, where the CHAOS data explicitly reports \ha, \nii, and \sii\ fluxes individually, instead of correcting our \ha\ flux for \nii\ and \sii, we compare our \emph{uncorrected} \ha\ fluxes to the combined CHAOS fluxes for the lines, i.e., $\mathrm{H}\alpha + \text{[N \textsc{ii}]} - \text{[S \textsc{ii}]}$. Finally, since the CHAOS fluxes are corrected for internal reddening, we correct our fluxes for reddening following the procedure outlined in Section~\ref{corrections}. 

\begin{figure*}
\epsscale{1.2}
\plotone{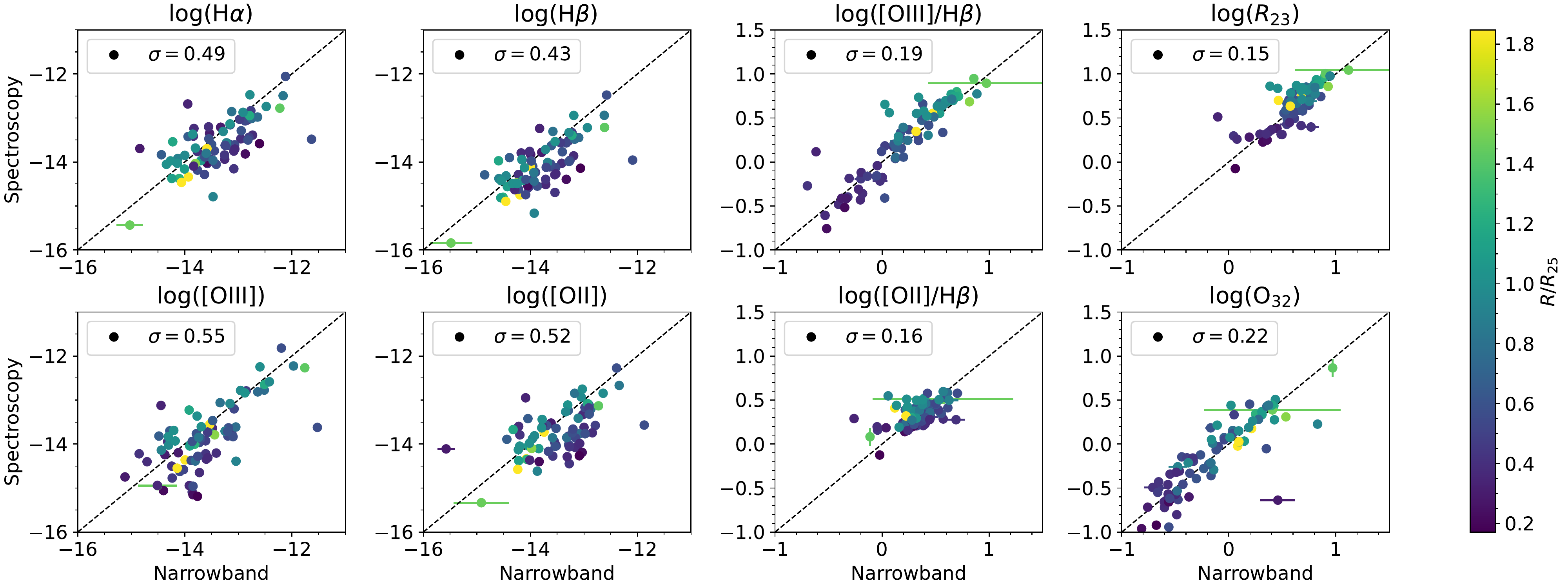}
\caption{From left to right, top to bottom: the \ha\ fluxes, \hb\ fluxes, \oiii/\hb\ ratio, $R_{23}$ ratio, \oiii\ fluxes, \oii\ fluxes, \oii/\hb\ ratio, and O$_{32}$ ratio for the 77 \hii\ regions in M101 selected by the CHAOS group. The fluxes are measured in \si{\erg\per\second\per\square\centi\metre}. All quantities shown are logarithmic. The points are colored by distance from the center of M101 in $R_{25}$. The dashed lines are $1:1$ lines, and the values reported in the legend are the scatter from the $1:1$ line, $\sigma$. The CHAOS spectroscopic flux is measured in \ang{;;1} slits, while our narrowband fluxes are measured in \ang{;;3} apertures, leading to the systematic offset seen in the line fluxes. The CHAOS \ha\ fluxes simulate our \ha\ flux measurements by adding the \nii\ fluxes and subtracting the \sii\ fluxes. \label{chaos_compare}}
\end{figure*}

Figure \ref{chaos_compare} shows the comparison of our photometric data with the spectroscopic data from CHAOS. Reported are each narrowband flux as well as four line ratios, \oiii/\hb, \oii/\hb, $R_{23}$ and O$_{32}$ (see Section \ref{sub:ratios}). The diagonal dashed lines in each plot is the $1:1$ line. Reported for each comparison is the scatter from the $1:1$ line, $\sigma$. Although there is some scatter in the line fluxes, the data is consistent with a linear trend. However, there is a slight vertical offset from the $1:1$ line in each of the line fluxes where CHAOS systemically measures lower line emission than in our photometry. This is likely caused by the effect mentioned above where our larger aperture size will measure more flux. 

\subsection{Strong-Line Ratios}\label{sub:ratios}

As mentioned in the introduction, the most direct and physically motivated method to determine the oxygen abundance of an \hii\ region or star-forming galaxy is to measure the electron temperature, $T_e$, of the ionized gas using the intensity of one or more temperature-sensitive auroral lines \citep{dinerstein1990,skillman1998,stasinska2007}. Unfortunately, these auroral lines are intrinsically faint, making them difficult to measure, and none of our filters target these lines. Therefore, we utilize strong-line abundance calibrations to estimate the metallicity of the ionized gas. The caveat is that strong-line methods are indirect and usually model dependent. 

Modern strong-line calibrations fall into two broad categories: empirical and theoretical. Empirical methods are calibrated against high-quality observations of individual \hii\ regions with measured direct (i.e., $T_e$-based) oxygen abundances (e.g., \citealt{pilyugin2000,pilyugin2001,pilyugin2005,peimbert2007}). However, these are limited as they are only strictly applicable to regions similar to those used to make the calibration and are lacking high-excitation \hii\ regions, especially in the metal-rich regime. Theoretical calibrations get around this issue by using photoionization model calculations across a wide range of nebular conditions (e.g., \citealt{mcgaugh1991,kewley2002,kobulnicky2004}). These are not without their problems as well, including oversimplified geometries and the unconstrained role that dust plays in the depletion of metals. 

The individual limitations of these various calibrations is compounded by the fact that there exists large, poorly understood systematic discrepancies, in the sense that strong-line abundances generally deviate from $T_e$-based abundances. Furthermore, empirical calibrations and theoretical calibrations disagree with each other. \cite{moustakas2010} found that empirical calibrations underestimate the ``true,'' $T_e$-based metallicity by $\sim$\SIrange{0.2}{0.3}{\dex}, while the theoretical calibrations yield abundances that are too high by the same amount. Consequently, the absolute uncertainty in the nebular abundance scale is a factor of $\sim$5 \citep{kewley2008}. Unfortunately, the physical origin of this systematic discrepancy remains unsolved. As such, when comparing trends between multiple calibrations, qualitative trends should be compared, not quantitative differences between the calibrations. 

While there are numerous strong-line calibrations in the literature with which to estimate abundances, our narrowband filters limit us to those calibrations that use the metallicity-sensitive $R_{23}$ parameter \citep{pagel1979}: 
\begin{equation}
	R_{23} = \frac{\text{[O \textsc{ii}]}\lambda3727 + \text{[O \textsc{iii}]}\lambda\lambda4959,5007}{\mathrm{H}\beta}.
\end{equation}
The advantage of $R_{23}$ as an oxygen abundance diagnostic is that it is directly proportional to both principal ionization states of oxygen, unlike other diagnostics that have a second-order dependence on the abundance of other elements like nitrogen or sulfur (e.g., \citealt{pettini2004,pilyugin2010}). The disadvantage of $R_{23}$ for our purposes is that it must be corrected for stellar absorption and dust attenuation. The most serious complication is that the relation between $R_{23}$ and metallicity is famously double-valued. Metal-rich objects lie on the upper $R_{23}$ branch, while metal-poor objects lie on the lower branch; the transition between the upper and lower branches happens around a metallicity of \SIrange{8.4}{8.5}{\dex} and is called the turn-around region. 

To add to the difficulty of using $R_{23}$ as an abundance indicator, there also exists a dependence on an ionization/excitation parameter. The ionization parameter, $q$, is defined as 
\begin{equation}
	q = \frac{S_{H^0}}{n},
\end{equation}
where $S_{H^0}$ is the ionizing photon flux per unit area and $n$ is the local number density of hydrogen atoms \citep{kewley2002}. The ionization parameter is usually measured through the ratio of the optical oxygen lines, 
\begin{equation}\label{eq:o32}
	\text{O}_{32} = \frac{\text{[\ion{O}{3}]}\lambda\lambda4959,5007}{\text{[\ion{O}{2}]}\lambda3727}.
\end{equation}
Some calibrations have attempted to take this into account \citep[e.g.][]{mcgaugh1991,kobulnicky2004}, while others have not \citep[e.g.][]{zaritsky1994}. The excitation parameter, $P$, was defined by \cite{pilyugin2000} in an analogous way to the ionization parameter: 
\begin{equation}\label{eq:p}
	P = \frac{\text{[\ion{O}{3}]}\lambda\lambda4959,5007}{\text{[\ion{O}{2}]}\lambda3727 + \text{[\ion{O}{3}]}\lambda\lambda4959,5007}.
\end{equation}

\begin{figure}
\epsscale{1.2}
\plotone{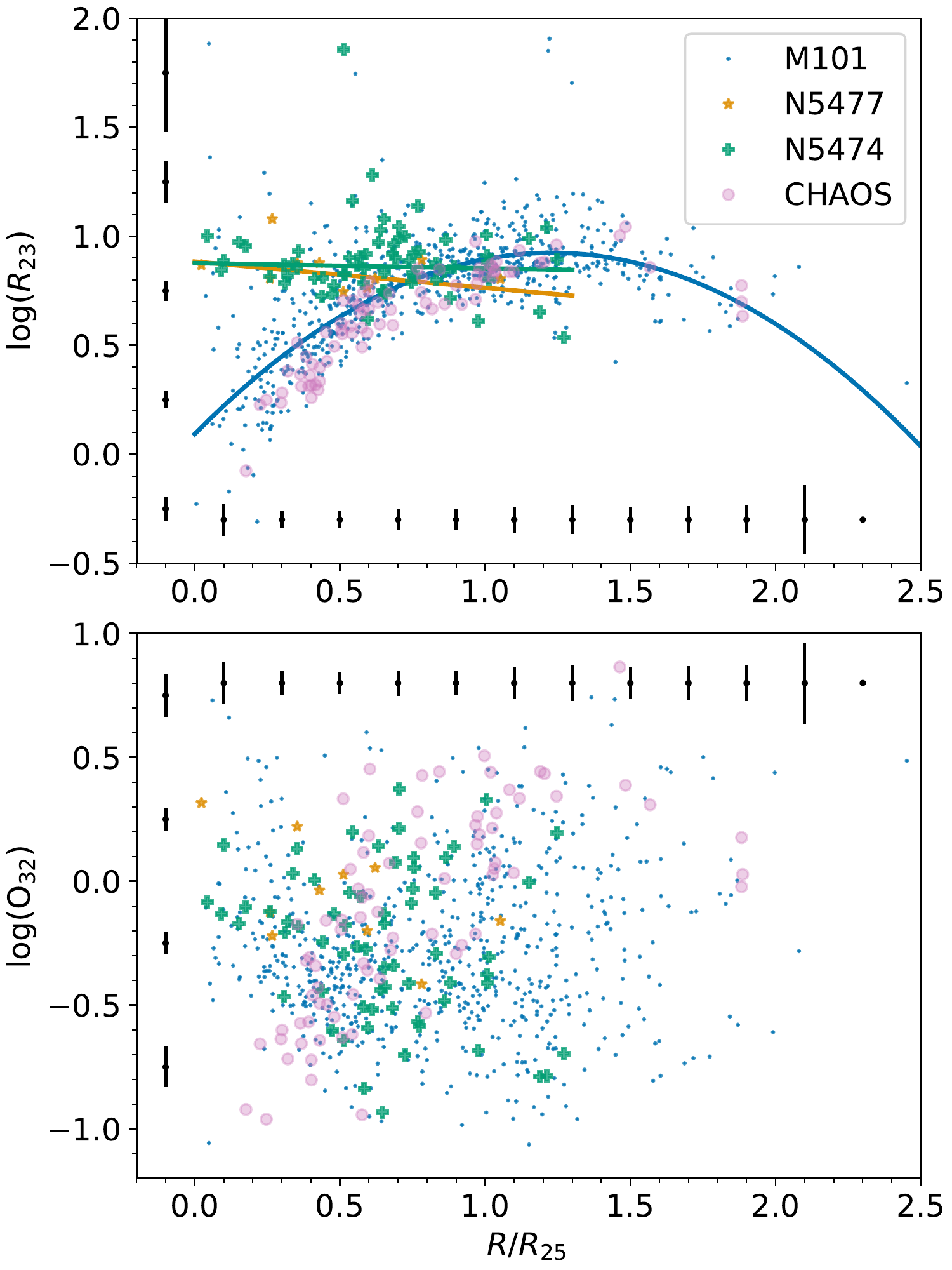}
\caption{Top: $\log(R_{23})$ as a function of radius using each galaxy's own $R_{25}$. Fitted lines are given to guide the eye to radial trends. Bottom: $\log(\mathrm{O}_{32})$ as a function of radius. In both plots, blue points are regions in M101 in our data, purple circles are regions in M101 from CHAOS, green plus signs are regions in NGC~5474, and orange stars are regions in NGC~5477. Black points with error bars show characteristic errors for both strong-line ratios as a function of radius and ratio. \label{strong_line}}
\end{figure}

Before we estimate oxygen abundances, we investigate the behavior of these strong-line ratios. Figure \ref{strong_line} shows the strong-line ratios $R_{23}$ and O$_{32}$ as a function of radius for each of the galaxies in our sample. For comparison, the spectroscopic data from CHAOS is plotted as well. The regions in M101 have an $R_{23}$ that strongly varies with radius: starting out low for the inner regions, reaching a peak at about $1.4R_{25}$, and decreasing further out. This likely reflects the non-monotonic relation between $R_{23}$ and metallicity, coupled with M101's radial metallicity gradient. This will aid in determining which branch a given \hii\ region is on when we estimate metallicities. Interestingly, the satellites do not have a strong radial gradient in $R_{23}$, which remains roughly constant throughout their populations, equivalent to the regions at intermediate radius of M101. 

The ionization state of the \hii\ regions in our sample, as measured by O$_{32}$, shows more scatter than does $R_{23}$. Noticeably, we measure more low ionization regions at larger radii in M101 than CHAOS does. The CHAOS data requires measurable auroral [\ion{O}{3}]$\lambda$4363, which limits that dataset to only the brightest regions at large radii. Since our detections are determined by measured stronger, collisionally-excited lines, we are able to measure fainter, and thus more \hii\ regions at any radius. Similar statements can be made about the satellites. 

\subsection{Abundance Calibrations \& Gradient Shape}\label{sub:bayesian}

As alluded to in Section \ref{sub:ratios}, there are poorly understood systematic discrepancies among strong-line calibrations: empirical calibrations generally yield oxygen abundances that are factors of $1.5$-5 times lower than abundances derived using theoretical calibrations \citep{kennicutt2003,garnett2004,bresolin2004,bresolin2005,shi2006,nagao2006,kewley2008,moustakas2010}. This being an unsolved issue, we have chosen to estimate our metallicities using three strong-line calibrations: the theoretical calibrations of \citet[][hereafter KK04]{kobulnicky2004} and \citet[][hereafter M91]{mcgaugh1991}; and the empirical calibration of \citet[][hereafter PT05]{pilyugin2005}. This allows us to bracket the range of oxygen abundances one would derive with other existing strong-line calibrations. 

All three of these calibrations rely on the $R_{23}$ parameter to estimate metallicities and thus have two solutions, one for the upper branch ($u$) and another for the lower branch ($\ell$). For the KK04 calibration, we have
\begin{align}\label{kk04:lower}
	12 &+ \log(\mathrm{O/H})_{\ell} = 9.40 + 4.65x - 3.17x^2 \nonumber \\
	&- \log(q)(0.272 + 0.547x - 0.513x^2),
\end{align}
and
\begin{align}\label{kk04:upper}
	12 &+ \log(\mathrm{O/H})_{u} \nonumber \\
	&= 9.72 - 0.777x - 0.951x^2 - 0.072x^3 \nonumber \\
	&\qquad -0.811x^4 - \log(q)(0.0737 - 0.0713x - 0.141x^2 \nonumber \\
	&\qquad +0.0373x^3 - 0.058x^4),
\end{align}
where $x = \log(R_{23})$. The ionization parameter $q$ in \si{\centi\metre\per\second} is given by
\begin{align}\label{kk04:ion}
	\log(q) &= 32.81 - 1.153y^2 + z(-3.396 - 0.025y + 0.1444y^2) \nonumber \\
	&\qquad \times [4.603 - 0.3119y - 0.163y^2 \nonumber \\
	&\qquad + z(-0.48 + 0.0271y + 0.02037y^2)]^{-1},
\end{align}
where $z = 12 + \log(\mathrm{O/H})$ and $y = \log(\mathrm{O}_{32})$, where O$_{32}$ is given by Equation \ref{eq:o32}. Note that Equations \ref{kk04:lower}-\ref{kk04:ion} must be solved iteratively for both the ionization parameter and the oxygen abundance; convergence is usually achieved in $\sim$3 iterations. 

For the M91 calibration, we use the functional form as parameterized by \cite{kuziodenaray2004}: 
\begin{align}\label{m91:lower}
	12 &+ \log(\mathrm{O/H})_{\ell} = 12 - 4.944 + 0.767x + 0.602x^2 \nonumber \\
	&\qquad -y(0.29 + 0.332x - 0.331x^2),
\end{align}
and
\begin{align}\label{m91:upper}
	12 &+ \log(\mathrm{O/H})_{u} = 12 - 2.939 - 0.2x - 0.237x^2 - 0.305x^3 \nonumber \\
	&\qquad - 0.0283x^4 - y(0.0047 - 0.0221x \nonumber \\
	&\qquad - 0.102x^2 - 0.0817x^3 - 0.00717x^4),
\end{align}
where $x = \log(R_{23})$ and $y=\log(\mathrm{O}_{32})$. 

Finally, for the PT05 calibration, we have
\begin{align}\label{pt05:lower}
	12 &+ \log(\mathrm{O/H})_{\ell} \nonumber  \\
	&= \frac{R_{23} + 106.4 + 106.8P - 3.40P^2}{17.72 + 6.60P + 6.95P^2 - 0.302R_{23}},
\end{align}
and
\begin{align}\label{pt05:upper}
	12 &+ \log(\mathrm{O/H})_{u} \nonumber \\
	&= \frac{R_{23} + 726.1 + 842.2P + 337.5P^2}{85.96 + 82.76P + 43.98P^2 + 1.793R_{23}},
\end{align}
where $P$ is given by Equation \ref{eq:p}.

Given these calibrations, we now ask what shape the oxygen abundance gradient should take. As mentioned in the introduction, either simple exponential models or broken exponential models are plausible descriptions for metallicity profiles in spiral galaxies. An investigation of Figure~\ref{strong_line} shows that the \hii\ regions in M101 have $R_{23}$ ratios that vary strongly with radius. The rapidly increasing $R_{23}$ values in M101's inner disk are well explained by the fact that inner regions of bright spiral galaxies have high metallicities and negative gradients \citep[e.g.][]{croxall2016}. However, beyond $1.3R_{25}$ the double-valued nature of the $R_{23}$-O/H relation makes it uncertain whether the data is better explained by a simple exponential model or a broken exponential model. Rather than assigning a branch and deriving metallicities, we use a Bayesian approach to investigate which model best describes the observed $R_{23}$ data.

In this Bayesian analysis, we first adopt an abundance model: a simple exponential with a gradient and zero point or a broken exponential with a gradient and zero point for the inner region of the galaxy and a break radius followed by a flat outer region. Additionally, both models have a parameter that gives the abundance scatter at a given radius, resulting in a total of three parameters for the simple exponential and four parameters for the broken exponential model. For a given model, we then adopt a calibration that maps oxygen abundance to $R_{23}$, i.e., the calibration of KK04. We adopt uniform priors on the values of each parameter, using the \texttt{Python} package \texttt{emcee} \citep{emcee} to solve for the fit parameters of each model. We report our priors and best fit parameters in Table \ref{bayesian_table}.

\begin{deluxetable}{l l l}
\tablecaption{Bayesian Priors and Fits \label{bayesian_table}}
\tablehead{\colhead{Model/Parameters} & \colhead{Priors} & \colhead{Best Fit}}
\startdata
\underline{Simple exponential} & & \\
Gradient & $U(-0.8, -0.3)$ \si{\dex\perradius} & $\num{-0.53}^{+0.02}_{-0.03}$ \si{\dex\perradius} \\
Zero Point & $U(8.7, 9.2)$ \si{\dex} & $+\num{9.13}^{+0.02}_{-0.01}$ \si{\dex} \\
Scatter & $U(0.0, 0.5)$ \si{\dex} & $+\num{0.02}^{+0.01}_{-0.01}$ \si{\dex} \\ \hline
\underline{Broken Exponential} & & \\
Gradient & $U(-0.8, -0.3)$ \si{\dex\perradius} & $\num{-0.56}^{+0.03}_{-0.04}$ \si{\dex\perradius} \\
Zero Point & $U(8.7, 9.2)$ \si{\dex} & $+\num{9.14}^{+0.02}_{-0.01}$ \si{\dex} \\
Break Radius & $U(0.7, 1.2)$ \si{\radius} & $+\num{1.06}^{+0.08}_{-0.08}$ \si{\radius} \\
Scatter & $U(0.0, 0.5)$ \si{\dex} & $+\num{0.03}^{+0.01}_{-0.01}$ \si{\dex} 
\enddata
\end{deluxetable}

\begin{figure*}
\plotone{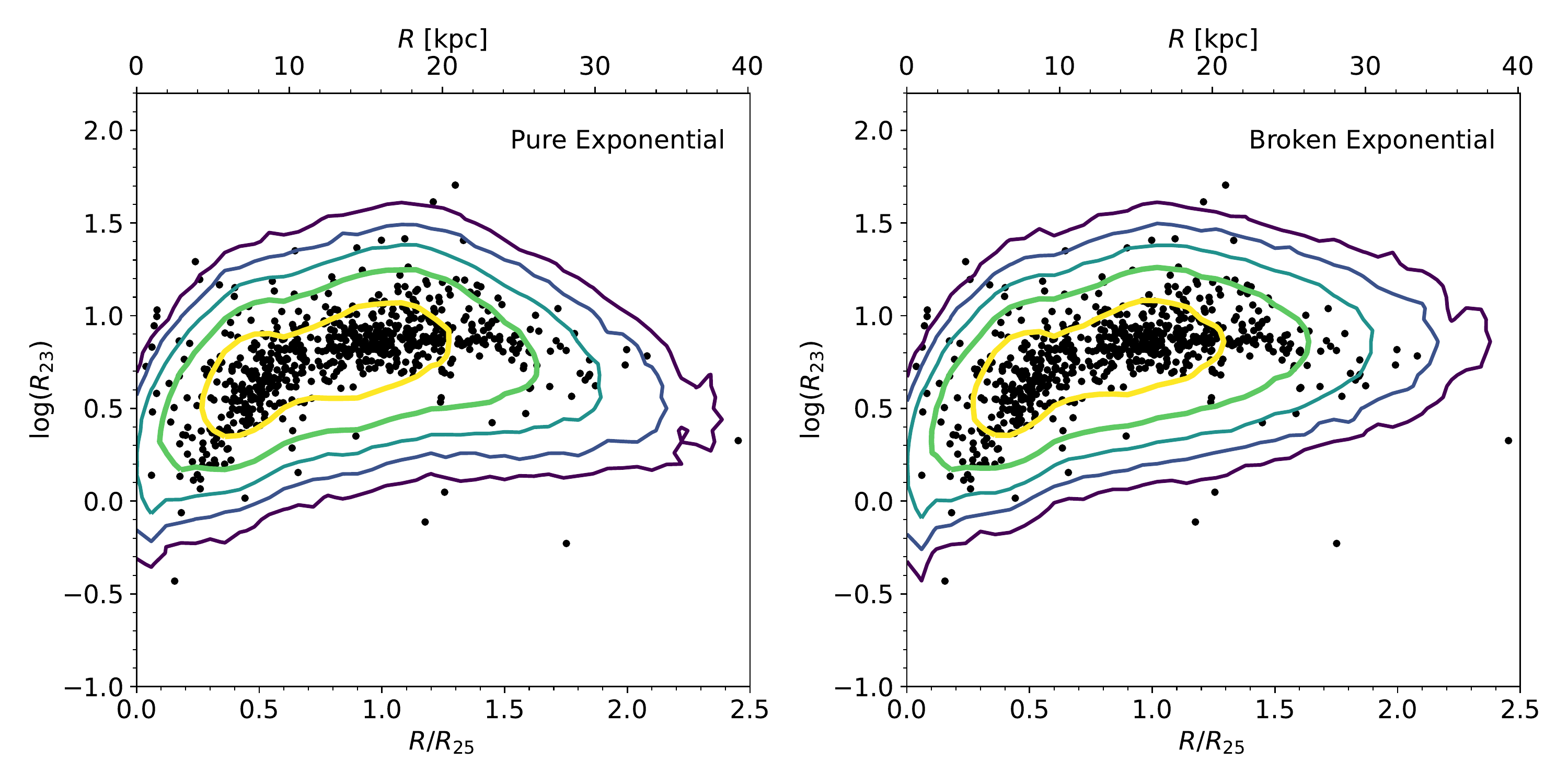}
\caption{Our measured $R_{23}$ data (black points) for M101 overlaid on the $R_{23}$ probability density contours for the simple exponential model (left) and broken exponential (right) generated using the best-fit Bayesian parameters described in the text. For both plots, the KK04 calibration was used to generate theoretical $R_{23}$ values. \label{model_r23}}
\end{figure*}

Using the best fit parameters from the Bayesian analysis, we generate probability density contours for the $R_{23}$ distribution in each model and overlay our observed data for M101 in Figure \ref{model_r23}. Comparing the two model contours to the data, we see that both models fit the $R_{23}$ data of the inner portion of M101 relatively well, but they differ in the outskirts. In the simple exponential model, the continuing decrease in metallicity results in probability contours which bend downwards in the outskirts, while in the broken model the flat abundances in the outskirts leads to a flattened $R_{23}$ profile over that same region. Neither of these trends are precisely described by the observed $R_{23}$ datapoints, leaving ambiguity as to the choice of the best model. Indeed, in reality, the proper model may lie in between these two extremes: a weak but still negative abundance gradient in the galaxy outskirts, similar to what has been found for other galaxies (M83, \citealt{bresolin2009}; NGC~4625, \citealt{goddard2011}; NGC~1058, \citealt{bresolin2019};).

For a more quantitative analysis, we calculated the \ac{aic} and \ac{bic} for both models (\citealt{akaike1974,schwarz1978}; but see \citealt{elements_of_stats} for an introduction). The \ac{aic} penalizes complex models less, while the \ac{bic} penalizes complex models more, so an agreement of the two quantities is a good statement of which model truly fits the data best. Both the \ac{aic} and \ac{bic} indicate that the simple exponential model fits the data slightly better than the broken exponential model. However, even with deep photometry, the lack of points at large radii makes the ``goodness of fit'' hard to pin down. This motivates a detailed spectroscopic follow up for the outer disk \hii\ regions of M101 to determine their $R_{23}$ ratios and oxygen abundances more accurately. In the following, we will calculate abundances for and discuss possible physical origins for both types of gradients in M101.

\subsection{Oxygen Abundances}\label{sub:oxy}

Given the calibrations in Section~\ref{sub:bayesian}, we now estimate the metallicities and reasonable uncertainties for each \hii\ region in our sample. As such, we follow the procedure outlined in \citet{moustakas2010}. The goal of this procedure is to estimate the oxygen abundances and uncertainties for regions that lie directly on the $R_{23}$-O/H relation as well as those that lie off of the relation but are statistically consistent with being on the relation given measurement uncertainties. We refer the reader to \citet[][in particular Fig.\ 6]{moustakas2010} for the details, but provide a brief summary here. 

For each \hii\ region in our sample, we calculate a metallicity for both the upper and lower branch solutions of a given calibration. We then repeat this for 500 trials in a Monte Carlo algorithm sampling within the (assumed) Gaussian errors on the \oii, \oiii, and \hb\ line fluxes. Comparing the histograms of the trials, there are three possible outcomes. If the central values of the two histograms are well-separated (i.e., beyond $1\sigma$ of each other and $(\text{O/H})_{\text{u}} > (\text{O/H})_\ell$), then the upper- and lower-branch abundances and their uncertainties are taken to be the median and standard deviation of each distribution. If $(\text{O/H})_{\text{u}} > (\text{O/H})_\ell$, but they are within $1\sigma$ of each other, we adopt the average of the two histograms as the oxygen abundance, with the standard deviation of the combined histograms as the associated uncertainty. These regions have abundances around the turn-around region; however, they also have large abundance errors, reflecting the $R_{23}$ branch uncertainty. Finally, if the upper- and lower-branch solutions are statistically inconsistent with one another, $(\text{O/H})_{\text{u}} < (\text{O/H})_\ell$, given the measurement uncertainties, then no solution exists and these objects are rejected. 

Having estimated metallicities and their uncertainties for each region and each calibration, we have to make a decision to which branch of the $R_{23}$-O/H relation each region belongs. A common criterion is to utilize the ratios \nii/\ha\ and \nii/\oii\ to break the degeneracy on a region-by-region basis \citep{contini2002,kewley2008}, but the \nii\ lines are unavailable to us. Instead, we can only make general statements about the populations as a whole. As we have seen in Section \ref{sub:bayesian}, there are two general shapes that the abundance gradient can take which requires two different assumptions for the regions in the outskirts.

Of the 746 regions, 10 regions, and 65 regions detected in M101, NGC~5477, and NGC~5474, respectively, 660/10/57 (\SI{88}{\percent}/\SI{100}{\percent}/\SI{88}{\percent}) regions had computed KK04 abundances, 679/10/57 (\SI{91}{\percent}/\SI{100}{\percent}/\SI{88}{\percent}) regions had computed M91 abundances, and 619/9/56 (\SI{93}{\percent}/\SI{90}{\percent}/\SI{86}{\percent}) had computed PT05 abundances. This gives us metallicity measurements in a total of 727/746/684 \hii\ regions scattered across the M101 Group for the KK04, M91, and PT05 calibrations, respectively. 

\subsection{The Gradients of NGC~5477 and NGC~5474}\label{sub:sat_grad}

\begin{figure}
\plotone{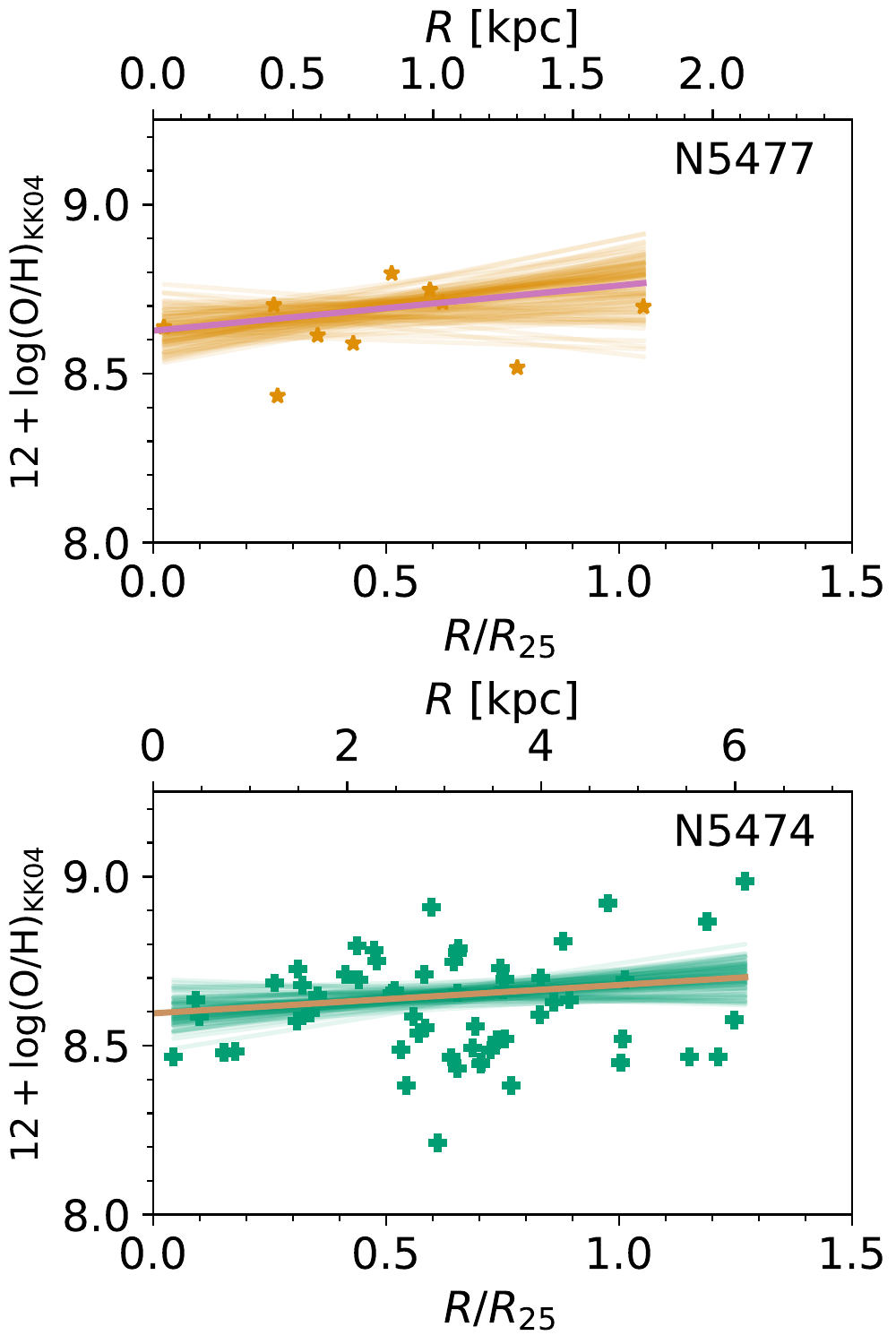}
\caption{KK04 metallicities and fitted radial gradients for the two satellites, NGC~5477 (top) and NGC~5474 (bottom). The solid purple and brown lines are the best-fit solutions, and the faded orange and green lines are 100 random walk fits that reflect the posterior distributions. Abundances are calculated via the upper branch of the $R_{23}$ relation (see text). \label{satellites}}
\end{figure}

Given the trends evident in Figure \ref{strong_line} for the two satellite galaxies, it is relatively straightforward to assign an $R_{23}$ branch to each galaxy's \hii\ regions. Both satellite galaxies' \hii\ regions have $R_{23}$ values consistent with being on the upper branch (see Figure \ref{strong_line}). We perform a Bayesian linear fit on each radial abundance gradient using the \texttt{Python} package \texttt{emcee} assuming flat priors and a Gaussian likelihood function. The radial abundance gradients are shown in Figure \ref{satellites} and the fit parameters are reported in Table \ref{exp_fits}. 

\begin{deluxetable}{lcccccc}
\tablecaption{Simple exponential Fits \label{exp_fits}} 
\tablehead{\colhead{Galaxy} & \multicolumn{2}{c}{Slope} & \colhead{Intercept} \\
\colhead{} & \colhead{dex$/R_{25}$} & \colhead{dex$/\si{\kilo\parsec}$} & \colhead{\si{\dex}}}
\startdata
\multicolumn{4}{l}{KK04:} \\
M101\tablenotemark{a} & \num{-0.53 \pm 0.01} & \num{-0.033 \pm 0.001} & \num{9.17 \pm 0.01} \\
NGC 5477 & \num{+0.14 \pm 0.07} & \num{+0.084 \pm 0.042} & \num{8.62 \pm 0.03} \\
NGC 5474 & \num{+0.08 \pm 0.04} & \num{+0.017 \pm 0.008} & \num{8.60 \pm 0.03} \\ \hline
\multicolumn{4}{l}{M91:} \\
M101\tablenotemark{a} & \num{-0.54 \pm 0.01} & \num{-0.034 \pm 0.000} & \num{9.02 \pm 0.01} \\
NGC 5477 & \num{+0.08 \pm 0.06} & \num{+0.048 \pm 0.036} & \num{8.53 \pm 0.03} \\
NGC 5474 & \num{+0.06 \pm 0.03} & \num{+0.012 \pm 0.006} & \num{8.47 \pm 0.02} \\ \hline
\multicolumn{4}{l}{PT05:} \\
M101\tablenotemark{a} & \num{-0.85 \pm 0.03} & \num{-0.053 \pm 0.002} & \num{8.89 \pm 0.02} \\
NGC 5477 & \num{-0.03 \pm 0.07} & \num{-0.018 \pm 0.042} & \num{8.29 \pm 0.03} \\
NGC 5474 & \num{+0.07 \pm 0.05} & \num{+0.015 \pm 0.010} & \num{8.17 \pm 0.03} 
\enddata
\tablenotetext{a}{Assumed that the $R_{23}$-O/H relation goes from upper to lower branch at $1.3R_{25}$ (see text).}
\end{deluxetable}

Both NGC~5477 and NGC~5474 have abundance gradients that are essentially flat regardless of the calibration given the small amount of regions dominated by scatter; the average abundance gradient for NGC~5477 and NGC~5474 are \SI{0.04 \pm 0.02}{\dex\per\kilo\parsec} and \SI{0.01 \pm 0.00}{\dex\per\kilo\parsec}, respectively. This is understandable given the small sizes of the satellites, $\sim$\SI{2}{\kilo\parsec} for NGC~5477 and $\sim$\SI{6}{\kilo\parsec} for NGC~5474, being that both galaxies are likely well-mixed. The relatively flat abundance gradient of NGC~5474 could be surprising given the likely interaction it had with M101 \citep[e.g.][]{waller1997,linden2022}. However, given the relatively small size of NGC~5474, gas mixing timescales over sub-kiloparsec scales being on the order of \SIrange{10}{100}{\mega\year} \citep{roy1995}, and the estimated age of the interaction being \SI{300}{\mega\year} \citep{mihos2018,linden2022}, it would be reasonable to assume that NGC~5474 had enough time to smooth out any large-scale variations. 

Despite the difficulty in investigating dwarf galaxies' oxygen abundances due to their small size and typically low luminosities, a few studies have been able to incorporate dwarfs into their analyses. The CALIFA survey analyzed galaxies down to a luminosity limit of $M_z < -17$ \citep{sanchez2014}, which includes galaxies like NGC~5474 but excludes NGC~5477. They found that a subsample of galaxies that are undergoing interactions have shallower abundance gradients, regardless of the stage of interaction, and independent of any other quality (morphology, mass, etc.). Those galaxies had an average gradient of $\sim-\SI[multi-part-units=single]{0.025 \pm 0.035}{\dex\perradius}$ \citep{sanchez2014}, consistent with the gradients we measure for these two satellites. This seems to support the scenario that NGC~5474 has interacted with M101.

However, caution must be taken when over-interpreting these trends. While it does appear that interacting galaxies do tend to have flatter abundance gradients than non-interacting galaxies \citep[e.g.][]{kewley2010,rich2012,sanchez2014}, this is not as striking a difference in dwarf galaxies. Numerous studies have found that regardless of the abundance estimator used, there is strong scatter within dwarf galaxies but a lack of any radial abundance gradient within the uncertainties of the abundance estimator \citep[e.g.][but see \citealt{pilyugin2015} for an alternative view]{pagel1978,roy1996,hunter1999,kniazev2005,vanzee2006,lee2007}. Therefore it is more likely that NGC~5474 and NGC~5477 have abundance gradients consistent with what is expected for dwarf galaxies.

\subsection{The Gradient(s) of M101}\label{sub:m101_grad}

\begin{figure*}
\plotone{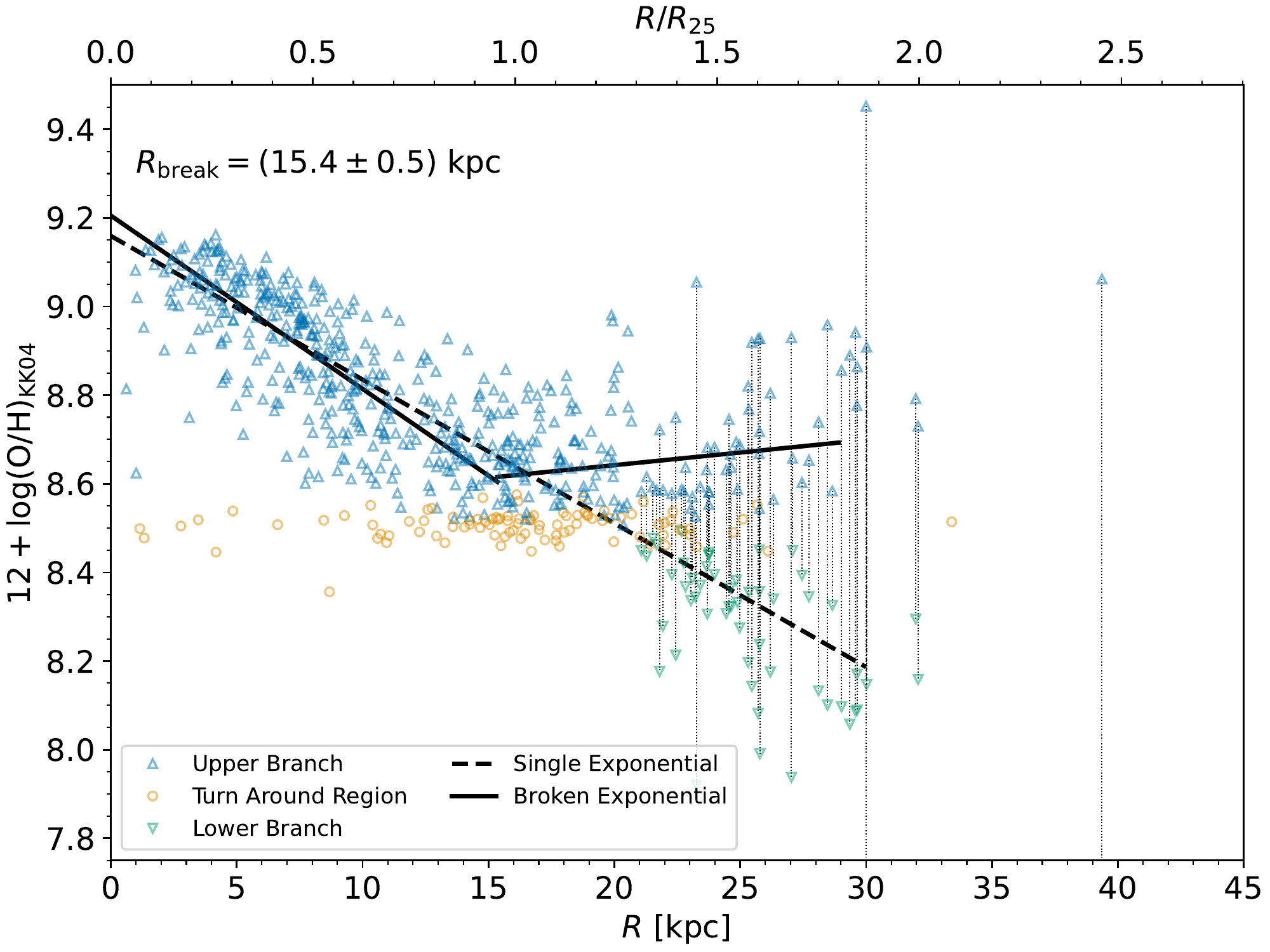}
\caption{The KK04 oxygen abundances of M101 fitted with a simple exponential (solid black line) and a broken exponential line (dashed black line). Regions in the turn-around region (orange circles), upper branch (blue triangles), and lower branch (green inverted triangles) are distinguished. Outside $1.3R_{25}$, we display both the upper and lower branch solutions connected by a dotted line for each individual \hii\ region. Characteristic error bars are $\pm\SI{0.02}{\dex}$ for the upper branch, $\pm\SI{0.04}{\dex}$ for the lower branch, and $\pm\SI{0.2}{\dex}$ for the turn-around region. As indicated in the plot, a break in the abundance gradient was detected at \SI{15.4 \pm 0.5}{\kilo\parsec}. \label{grad_break}}
\end{figure*}

While the satellite galaxies' abundance determinations was relatively straightforward, we now turn to the more complex situation of M101, where we consider both the simple and broken exponential models. Figure \ref{grad_break} shows the abundance gradient for M101. Beyond $1.3R_{25}$, both upper and lower branch solutions are shown for each individual \hii\ region connected by a dotted line to represent the two general abundance shapes assumed. If we assume that those \hii\ regions are located on the more metal-poor, lower branch, then we recover a simple exponential gradient across all calibrations consistent with spectroscopic studies (see Table \ref{exp_fits} for the fit parameters). For comparison, CHAOS reported a gradient of $-\SI[multi-part-units=single]{0.462\pm0.024}{\dex\perradius}$ ($-\SI[multi-part-units=single]{0.029 \pm 0.001}{\dex\per\kilo\parsec}$), corrected for our difference in $R_{25}$ and distance to M101. Adopting this simple exponential fit would imply that M101's abundance gradient is consistent with a simple application of the inside-out galaxy growth scenario. 

To recover the broken exponential shape, we assign the outer disk \hii\ regions to the more metal-rich, upper branch. As mentioned before, this decision is motivated by the radial trend of $R_{23}$ for M101 shown in Figure~\ref{strong_line} and based on evidence from other galaxies in the literature (see Section~\ref{sec:intro}). To identify the location of the break in the radial abundance gradient of M101, we fit the abundances with two exponential fits joined by a break (a segmented linear regression). The break position was estimated iteratively following the method outlined in \citet{muggeo2003} and implemented by the \texttt{Python} package \texttt{piecewise-regression} \citep{piecewise}. Starting with a model of a line with a term that incorporates a change in gradient between two segments of a piecewise function and given an initial guess of the break position, this model is fitted to the data with ordinary linear regression and bootstrap restarting \citep{wood2001} is used to find a new break position. The process is iterated until the position of the break converges. In this way an estimate on the uncertainty of the break position is also returned. This process is conceptually similar to that used in \citet{scarano2013}.

\begin{deluxetable*}{lccccc}
\tablecaption{M101 Broken Exponential Fits \label{brk_fits}}
\tablehead{\colhead{Calibration/Region} & \multicolumn{2}{c}{Slope} & \colhead{Intercept} & \multicolumn{2}{c}{Radial Break} \\
\colhead{} & \colhead{$\si{\dex}/R_{25}$} & \colhead{\si{\dex}/\si{\kilo\parsec}} & \colhead{\si{\dex}} & \colhead{$R/R_{25}$} & \colhead{\si{\kilo\parsec}}} 
\startdata
KK04/Inner & \num{-0.63 \pm 0.02} & \num{-0.039 \pm 0.001} & \num{9.21 \pm 0.01} & \multirow{2}{*}{\num{0.96 \pm 0.03}} & \multirow{2}{*}{\num{15.4 \pm 0.5}} \\
KK04/Outer & \num{+0.09 \pm 0.03} & \num{+0.006 \pm 0.002} & \num{8.53 \pm 0.03}  \\
M91/Inner & \num{-0.63 \pm 0.02} & \num{-0.039 \pm 0.001} & \num{9.08 \pm 0.01} & \multirow{2}{*}{\num{0.89 \pm 0.03}} & \multirow{2}{*}{\num{14.4 \pm 0.4}}  \\
M91/Outer & \num{+0.06 \pm 0.02} & \num{+0.004 \pm 0.001} & \num{8.43 \pm 0.02}  \\
PT05/Inner\tablenotemark{a} & \num{-0.85 \pm 0.03} & \num{-0.053 \pm 0.002} & \num{8.89 \pm 0.02} & \multirow{2}{*}{\num{0.65 \pm 0.04}} & \multirow{2}{*}{\num{10.5 \pm 0.7}} \\
PT05/Outer\tablenotemark{a} & \num{+0.08 \pm 0.02} & \num{+0.005 \pm 0.001} & \num{8.15 \pm 0.03}  
\enddata
\tablenotetext{a}{The fitted data was limited to only those regions for which this calibration strictly applies, i.e., $P \geq 0.4$.}
\tablecomments{The inner/outer distinction refers to the data fitted inside and outside the breakpoint listed in the last column.}
\end{deluxetable*}

Figure~\ref{grad_break} shows the result of this process for the KK04 calibration and Table \ref{brk_fits} records the results of the fits to all of the calibrations. We find a radial break in the KK04 and M91 calibrations at \SI[multi-part-units=single]{0.96 \pm 0.03}{\radius} and \SI[multi-part-units=single]{0.89 \pm 0.03}{\radius} (or \SI[multi-part-units=single]{15.4 \pm 0.5}{\kilo\parsec} and \SI[multi-part-units=single]{14.4 \pm 0.4}{\kilo\parsec}), respectively. We further distinguish between the ``inner'' and ``outer'' portions of M101 as defined by those \hii\ regions lying within that calibration's radial break and those lying without. While the fitted gradient in the outskirts of M101 shows a slightly positive slope, given both the large scatter in derived abundances and the large uncertainties in abundances near the $R_{23}$ turn-around region, this is likely more consistent with a simple flattening of the gradient.

\begin{figure}
\plotone{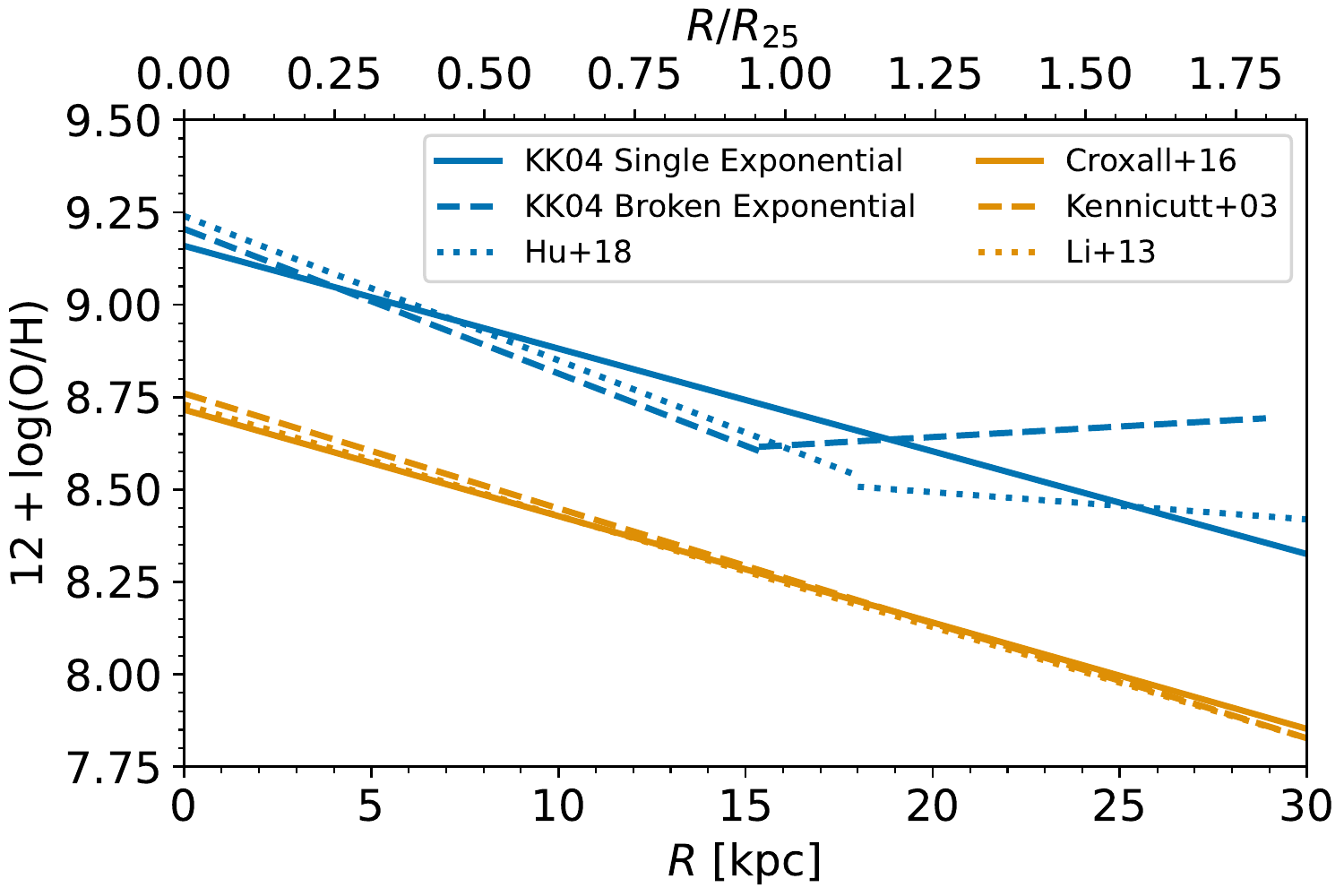}
\caption{A comparison of strong-line (blue lines) and $T_e$-based (orange lines) abundances for M101. Included are our simple and broken exponential fits for the KK04 abundance estimator, as well as the strong-line abundance gradient from \citet{hu2018} who also used the KK04 abundance estimator. The $T_e$-based abundance gradients are from \citet{croxall2016}, \citet{li2013}, and \citet{kennicutt2003}. In all cases from the literature, their gradients have been converted to our choice of $R_{25}$ and our assumed distance. \label{grad_compare}}
\end{figure}

Figure~\ref{grad_compare} compares our fitted gradients using the KK04 abundances to other abundance gradients found in the literature. As mentioned in Section~\ref{sub:ratios}, there is an unexplained deviation between $T_e$-based abundances and strong-line abundances of $\sim$\SI{0.5}{\dex} \citep[e.g.][]{kewley2008,moustakas2010}. Therefore, the discrepancy between the zero points of the spectroscopic abundances (orange lines) and the strong-line abundances (blue lines) is explained by that difference inherent to abundance estimations. Correcting for our choice of $R_{25}$ and our assumed distance, the slope of our simple exponential and the slope of the inner portion of the broken exponential is slightly steeper than that of the spectroscopic studies \citep{kennicutt2003,li2013,croxall2016}, but consistent with the other strong-line abundances \citep{hu2018}. The position of the radial break is also consistent with previous studies ($\sim$\SI{18}{\kilo\parsec}; \citealt{hu2018}).

For the PT05 calibration, we note that this calibration is empirical and can strictly only be applied to regions that are similar to those that were used to make the calibration. For PT05, that restricts \hii\ regions to those with $P \geq 0.4$ \citep{pilyugin2005,moustakas2010}. Implementing this restriction to our data results in a good fit and a break location that is consistent with the theoretical calibrations of \SI[multi-part-units=single]{0.65 \pm 0.04}{\radius} (\SI[multi-part-units=single]{10.5 \pm 0.7}{\kilo\parsec}). The fit with the constraint on the excitation parameter is what is reported in Table \ref{brk_fits}. We will explore the physical meaning of the break in Section \ref{sec:discussion}. 

\section{Azimuthal Asymmetries}\label{sec:azi}

As our data completely samples the disk of M101, an important test to undertake is to search for any large-scale azimuthal variations. The fundamental processes that govern chemical enrichment in a galaxy operate on different scales. For instance, oxygen is first produced in high-mass stars and then blown to parsec scales through winds or supernovae and then dispersed to kiloparsec scales by various mixing processes \citep{roy1995}. These should be represented in azimuthal deviations from a simple radial gradient since the timescale for galactic differential rotation to homogenize the interstellar medium ($\lesssim\SI{1}{\giga\year}$) is longer than both the oxygen production timescale ($<\SI{10}{\mega\year}$) and mixing timescales (\SIrange{10}{100}{\mega\year}; \citealt{roy1995}). 

\begin{figure*}
\plotone{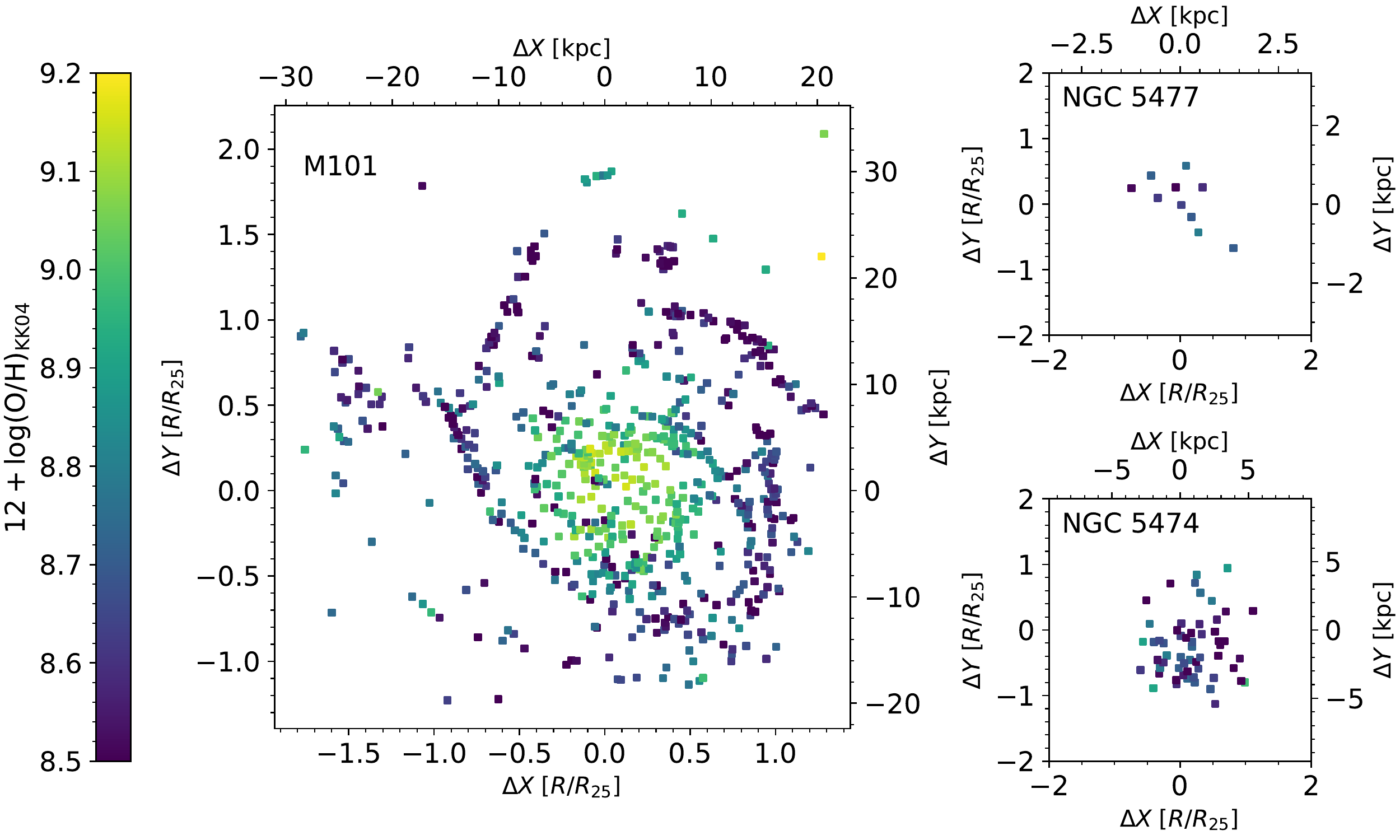}
\caption{The spatial distribution of KK04 abundances for M101 (left), NGC~5477 (top right), and NGC~5474 (bottom right). Points are colored by KK04 abundance (color bar to the left), and angular and physical units are given. \label{map}}
\end{figure*}

In Figure \ref{map}, we plot the spatial distribution of \hii\ regions in M101, NGC~5477, and NGC~5474. Each region is colored by its KK04 metallicity. An inspection of Figure \ref{map} shows no clear morphology to the abundance patterns beyond a radial gradient in all three galaxies. The lack of azimuthal abundance variations in the satellite galaxies is likely evidence of galactic homogenization for the same reasons as for their flat radial abundance gradients (Section \ref{sub:sat_grad}). There are minor variations along the spiral arms of M101, possibly similar in nature to the detached outer arm at $\sim$$R_{25}$ (position angles \ang{230} to \ang{290}) referred to as ``arc A'' by \citet{li2013}. Relative to the abundance scatter outside of ``arc A,'' the abundance scatter they detected is at the $2\sigma$ level. Additionally, there does not appear to be any major difference between \hii\ regions lying in a spiral arm and those lying between arms. 

As a first test towards a quantitative understanding of azimuthal variations, we investigate the reported asymmetry in abundances of M101 by \citet{kennicutt1996}. They suggested that \hii\ regions in the SE have lower oxygen abundances compared to \hii\ regions in the NW as obtained from the use of the $R_{23}$ parameter. Later studies reported that there did not appear to be a large-scale azimuthal difference between the SE and NW halves of M101 \citep{kennicutt2003,li2013}. To test this with our expanded sample of \hii\ regions, we divided M101 into two halves with respect to the major axis: a SE part (position angles \ang{37} to \ang{217}, including 270 regions) and a NW part (complementary position angles, with 390 regions). We investigated the radial dependence of $R_{23}$ as well as the KK04 abundances and found that for both quantities, the distribution of data points is virtually the same between the two subsamples. 

\begin{figure}
\plotone{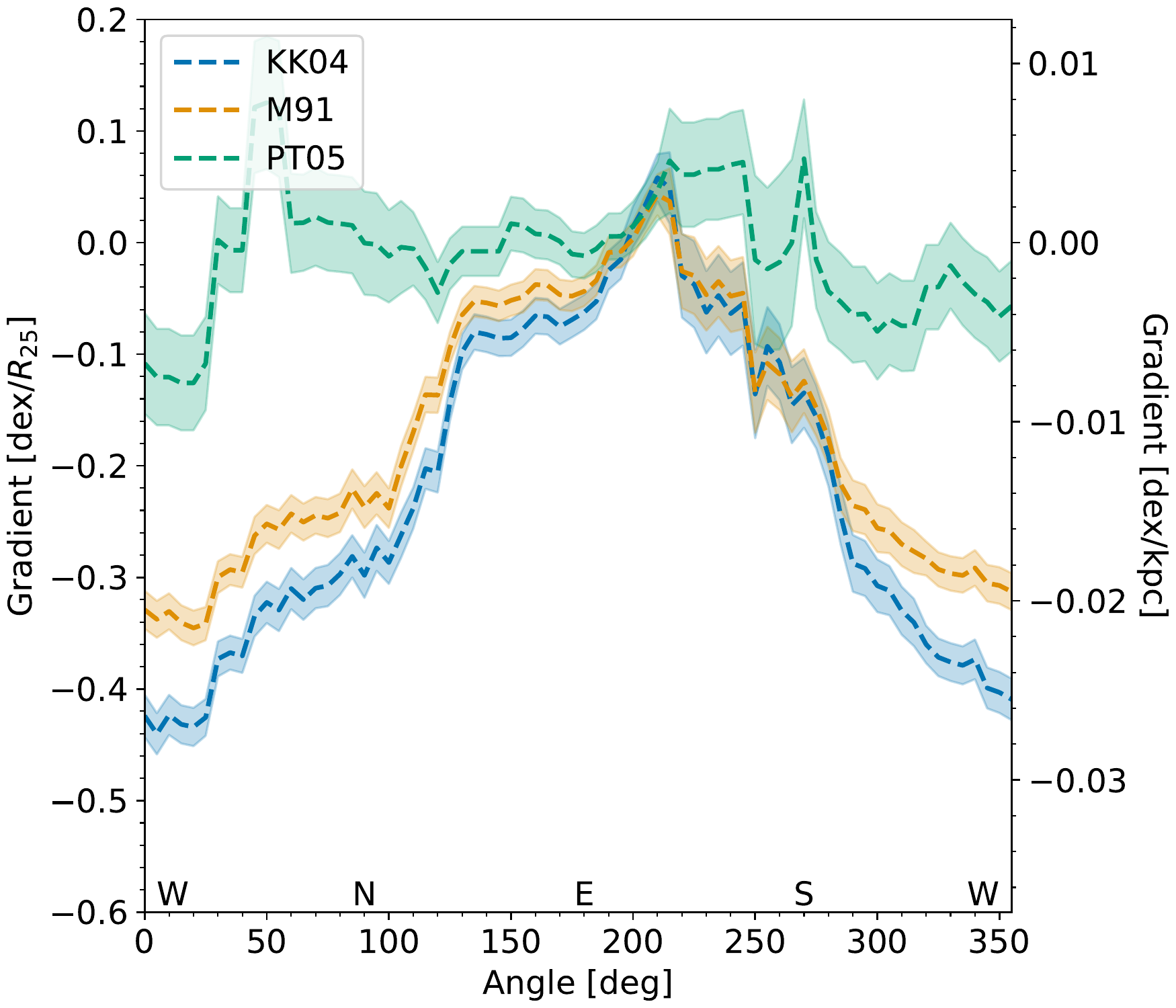}
\caption{The gradient of the outer regions of M101 in a quadrant rotated by \ang{5}. Only those \hii\ regions that satisfy $0.5 < R/R_{25} < 1.8$ were used in calculating gradients. The calibrations are as shown in the legend.  \label{quad}}
\end{figure}

However, given M101's strongly distorted disk, it is possible that there exists azimuthal differences on smaller angular scales or even between different halves of the galaxy. To test this, we measure the abundance gradient in M101's outer disk using azimuthal quadrants advanced incrementally by \ang{5}. The results for each calibration are plotted in Figure \ref{quad}. We see that regardless of the calibration, the western and southwestern regions of the galaxy have significantly steeper slopes than the rest of the galaxy, a difference of about \SI{0.4}{\dex}. As a further test, we varied the size of the wedge used, ranging from halves to octants, and found the same general trends. If there does exist an azimuthal bifurcation in M101, Figure \ref{quad} seems to suggest that the split is almost east-west rather than SE/NW. 

\section{Discussion}\label{sec:discussion}

In the previous sections, we have estimated abundances for $\sim$700 \hii\ regions throughout the M101 Group. When fitting a profile to the radial abundance gradient of M101, we find ambiguity between two possible models; both a simple exponential and a broken exponential gradient fits the data reasonably well depending on the method used. Additionally, we have also presented evidence of azimuthal abundance variations between the northeastern and southwestern halves of M101. In the following, we will explore these results in the context of M101's asymmetry and interaction history. 

\subsection{Galaxy Interactions \& the M101 Group}\label{sub:interaction}

Galaxy interactions and mergers are fundamental to galaxy formation and evolution and likely lead to strong mixing of abundances in the gas-phase metallicity. As laid out by \citet{toomre1972} and summarized in \citet{toomre1977}, two interacting galaxies will tidally interact during even a minor flyby. Tidal interactions and their associated shocks are thought to trigger gas flows throughout the disk \citep[e.g.][]{bushouse1987,kennicutt1987,barnes1996,mihos1996}. These gas flows should not only cause bursts of star formation, but it is reasonable to assume that radial mixing of metals in a galaxy should occur as well \citep[e.g.][]{lacey1985,schonrich2009,bresolin2009,rupke2010,bird2012}. Therefore, any pre-interaction gas-phase metallicity that exists should effectively be averaged out and imbued with new metals from post-interaction starbursts or accreted gas from the companion galaxy.

\begin{figure*}
\plotone{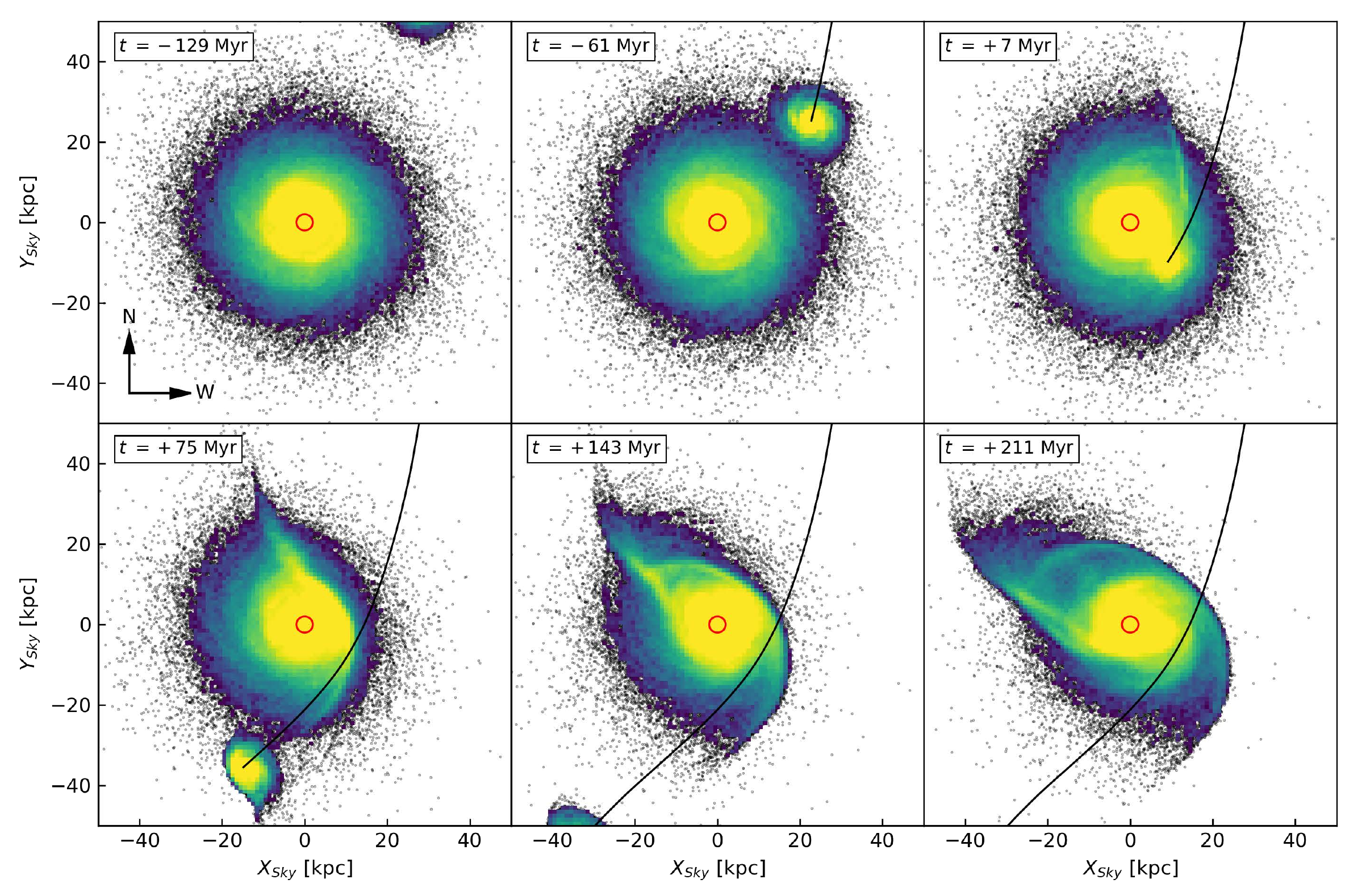}
\caption{The M101-NGC~5474 interaction from the Linden model. Time progresses from left to right, top to bottom. The orbit of NGC~5474 is traced in black and the stars in both galaxies are colored by local density of points. The red circle is of radius \SI{2}{\kilo\parsec} centered on the center of M101. Negative time indicates time before closest approach ($\sim$\SI{14}{\kilo\parsec}) and the last panel (bottom right) shows the current day arrangement of the galaxies. \label{sim}}
\end{figure*}

An example of where these mixing processes as a result of a flyby occurred is found in the M101 Group, where evidence of a flyby abounds throughout the group. The asymmetric disk of M101 has long been believed to have arisen from an interaction \citep{beale1969,rownd1994,waller1997}. In recent years, \ac{LSB} tidal features have been detected in the outskirts of M101 \citep{mihos2013} with colors and stellar populations consistent with a burst of star formation $\sim$\SIrange{300}{400}{\mega\year} ago \citep{mihos2018}. \citet[][hereafter the Linden model]{linden2022} recently modeled the system with a grazing retrograde encounter between M101 and NGC~5474 with closest approach occurring roughly \SI{200}{\mega\year} ago. Their simulation accurately reproduces the tidal morphology of M101 and shows the large radial motion imprinted in the galaxy's disk. Of particular importance, the simulation shows that the young starburst population seen in the NE Plume came from star formation triggered at the contact point when the galaxies collided.

Given the strong observational evidence cited above for an interaction between M101 and NGC~5474 joined with the Linden model, what does this mean for the shape of the radial abundance gradient of M101? Since we have two possible models for the abundance gradient of M101, we will investigate physical mechanisms for producing each gradient.

\subsection{A Simple Exponential Gradient\dots}

A simple exponential abundance gradient is predicted by the inside-out growth scenario for disk galaxies \citep[e.g.][]{scannapieco2009}, and it is likely that M101 followed suit. Since the metallicity of the stars and gas in a galaxy are decoupled, a hint of the original abundance gradient should be present in the old stellar populations. Using \emph{HST} photometry, \citet{mihos2018} were able to estimate the metallicities of old RGB stars in the NE Plume and stellar halo at projected distances of \SI{36}{\kilo\parsec} and \SI{47}{\kilo\parsec}, respectively. Assuming a solar oxygen abundance of $12 + \log(\text{O/H}) = 8.69$ \citep{asplund2009}, their metallicities correspond to an oxygen abundance of \num{7.39} and \num{6.99} for the NE Plume and halo, respectively. If instead of measuring the stellar halo, they measured the extreme outer disk, the slope between these two stellar measurements ($-\SI{0.036}{\dex\per\kilo\parsec}$) is consistent with the slopes of the inner disk found with the strong-line calibrations, hinting that the original abundance gradient of M101 was a simple exponential gradient as expected. 

However M101, like most galaxies, has not evolved in isolation. Although the M101 Group has been called one of the poorest groups in the Local Volume \citep{bremnes1999}, it being dominated by M101 with a scattering of low-mass companions, its galaxies likely have interacted with each other. As described in Section~\ref{sub:interaction}, M101 and its most massive companion, NGC~5474, likely interacted between \SIrange{200}{400}{\mega\year} ago \citep{mihos2018,linden2022}. This interaction was strong enough to influence the shape of M101's disk, but was it strong enough to influence its oxygen abundance? 

Numerous studies in the literature have investigated the integrated gas-phase metallicity of local galaxies and whether there is a dependence on environment \citep[e.g.][]{shields1991,skillman1996,mouhcine2007,ellison2009,kewley2010,peng2014}. While galaxies that reside in denser environments are found to have higher metallicities, this effect is small; the abundance difference between galaxies in dense and underdense regions is $<$\SI{0.1}{\dex} \citep{peng2014,pilyugin2017_env}. Relatively little attention has been paid to the abundance distribution within galaxies, i.e., their gradients, and any environmental dependence. \citet{lian2019} found that there was no change to the shape of the abundance gradient with respect to environment for massive galaxies which confirms earlier work \citep{kewley2006,kacprzak2015,pilyugin2017_env}. It seems that even in a strongly interacting environment like a galaxy group or cluster, the abundance profiles of galaxies do not deviate strongly from the simple exponential model. Perhaps M101 would still retain its original simple exponential abundance gradient despite the interaction with NGC~5474. 

However we must express caution at accepting that conclusion for an asymmetric galaxy like M101. The strong asymmetries present in M101's disk will cause its properties to vary as a function of azimuth. For example, \citet{mihos2013} constructed surface brightness profiles for M101, both azimuthally averaged and in octants. While the azimuthally averaged profile was a Type I profile \citep{pohlen2006}, i.e., fitted by a simple exponential gradient, the octants revealed the asymmetry. The northeastern portion of the disk was consistent with a Type III (upbending) profile and the southwestern portion of the disk was consistent with a Type II (downbending) profile. 

Not only are the two halves with different surface brightness profiles the same halves that have different radial abundance slopes (Figure~\ref{quad}) in M101, but this link between the abundance gradient and surface brightness profile is found in other galaxies as well \citep[e.g.][]{webster1983,edmunds1984,vilacostas1992,sanchez2014,pilyugin2014,bresolin2015}. Importantly, Type III surface brightness profiles, such as that found in the northeastern portion of M101's disk, are connected with abundance flattening at large radii \citep{marino2016}. Galaxies with such profiles are thought to have experienced episodes of inside-out growth in their outer disks in the past \citep{bresolin2012,marino2016} likely caused by tidal interactions with nearby galaxies \citep{watkins2019}. 

\subsection{\dots Or A Broken Gradient?}

Given the asymmetries present in M101's disk, let us now investigate the other abundance profile presented in the current study: the broken exponential gradient. As mentioned before (Section~\ref{sub:bayesian}), abundance profiles with breaks at $\sim$$R_{25}$ beyond which the abundances plateaus have been found in many galaxies \citep[e.g.][]{bresolin2009,goddard2011,bresolin2012,sanchez2014,sanchezmenguiano2016}. However, the physical nature of this flattening is still an open question since the ability to measure the metal content of the outermost regions of spiral galaxies is relatively recent. Numerous physical processes have been proposed, ranging from radial gas flows \citep[e.g.][]{lacey1985,goetz1992,schonrich2009}, resonance scattering with spiral density waves \citep{sellwood2002}, perturbations by satellite galaxies \citep{quillen2009,bird2012}, or highly efficient star formation in outer disks \citep{esteban2013}. Likely it will be some combination of these effects that produce the observed flattening at large radii.

In the specific case of M101, we believe that the interaction with NGC~5474 created a burst of star formation in the outer disk of M101 \citep{mihos2013,mihos2018}. The burst would have imbued the \ac{ISM} with oxygen and other metals created in the cores of short-lived massive stars, having the effect of raising the oxygen abundance in M101's disk. However, if the induced gas flows from the interaction were radially restricted, then instead of the entire disk rising in abundance, only the outer disk would increase in abundance, resulting in a flattening of the radial abundance gradient. We propose that this barrier is caused by the corotation radius. As predicted by theory \citep{mishurov2009} and seen in simulations \citep{lepine2001} and observations \citep[e.g.][]{marochnik1972,mishurov1999,dias2005,amores2009,elmegreen2009,lepine2011}, corotation has a ``pumping out'' effect: the corotation barrier produces gas flows in opposite directions on the two sides of corotation, with radial gas flows inward inside corotation and outward on the other side. 

\citet{scarano2013} compiled results from the literature of the corotation radii and metallicity breaks for 16 galaxies and found a strong correlation. Using the M101 rotation curve from \citet{roberts1975}, \citet{scarano2013} derive a corotation radius for M101 of \SI{15.6 \pm 2.2}{\kilo\parsec}, adjusted for our assumed M101 distance. This is consistent with the breaks in radial abundance gradients we have found for all three metallicity calibrations we used. Therefore, it is likely that the corotation of M101 is responsible for creating and maintaining the abundance gradient break after the interaction with NGC~5474. 

Important to this theory is the longevity of the corotation radius. If the corotation radius was not stationary on long timescales, i.e., if spiral structure was transient \citep{sellwood1984,sellwood2011}, then any break in radial abundance would be smoothed out. \citet{lepine2011} estimated the minimum lifetime of the Galactic corotation to be about \SI{3}{\giga\year}. Based on the same argument, this can be considered an estimate for the lifetime of corotation for M101 as well \citep{scarano2013}. This timescale could give the inner and outer disks of M101 enough time to chemically evolve differently before and after the interaction with NGC~5474. However, it should be mentioned that the dynamics of the interaction could reasonably change the stationary nature of corotation on short timescales. 

Once again, we must express caution at readily accepting this scenario given our data analysis techniques. \citet{pilyugin2003} noted that abundance breaks found using the $R_{23}$ strong-line method may not be real. This is caused by a systematic error depending on the excitation parameter $P$ (Eq.~\ref{eq:p}) where the method overestimates oxygen abundances in low excitation regions \citep{kinkel1994,castellanos2002,pilyugin2003}. To combat this excitation-dependence, we used the so-called $P$-method presented by \citet{pilyugin2003} and still found a radial abundance break at \SI[multi-part-units=single]{0.65 \pm 0.04}{\radius} (\SI[multi-part-units=single]{10.5 \pm 0.7}{\kilo\parsec}), albeit more interior than the breaks found with the M91 and KK04 calibrations.


\section{Conclusions}

We have conducted a narrowband emission-line survey to measure the oxygen abundance of the entire M101 Group, including M101 and its two major satellites, NGC~5477 and NGC~5474. Using narrowband filters that target \ha, \hb, \oiii, and \oii, we have detected a total of $\sim$930 \hii\ regions across these three galaxies. Our multi-band detection scheme lets us reject contamination due to foreground and background objects. After correcting for stellar absorption features, \nii\ and \sii\ emission, and internal extinction, our data shows a good correspondence to spectroscopic flux measurements \citep{croxall2016}. Additionally, our sample extends down to fainter \hii\ regions in all three galaxies across a wide range of ionization states compared to spectroscopic surveys. 

We use our emission line data to derive \hii\ region oxygen abundances for the M101 Group using the $R_{23}$ ratio \citep{pagel1979}. Specifically, we estimated the oxygen abundances with three strong-line calibrations: two theoretical calibrations \citep{mcgaugh1991,kobulnicky2004} and one empirical calibration \citep{pilyugin2005}. While these methods have an inherent uncertainty due to the double-valued nature of the $R_{23}$-O/H relation at low abundances, we present a variety of models that span the range of abundance patterns consistent with the data. Our main conclusions are summarized below. 

\begin{enumerate}

\item Of 853 \hii\ regions detected in M101, 11 \hii\ regions detected in NGC~5477, and 71 \hii\ regions detected in NGC~5474, we measured oxygen abundances for roughly \SI{75}{\percent} of the \hii\ regions in the M101 Group giving us a total of $\sim$720 \hii\ regions to trace the abundance patterns in the M101 Group. 

\item While the two satellite galaxies' oxygen abundances were best fitted with a simple flat gradient, M101's metallicity gradient required more detailed analysis. We compared two models for M101's gradient, a simple exponential model motivated by spectroscopic analyses of M101 and a broken exponential model motivated by spectroscopic analyses of a number of galaxies. Both models provide a plausible description of the data, with some suggestion that an intermediate model of a radial break to a slightly shallower gradient in the outskirts rather than a true flattening might be warranted.

\item Quantitatively, the simple exponential model for M101 has an exponential decline of \SI{-0.03}{\dex\per\kilo\parsec}, while the satellites were consistent with having shallow or flat gradients within the uncertainties. The broken exponential model for M101 has an exponential decline of \SI{-0.04}{\dex\per\kilo\parsec} inside a break radius of \SI{14}{\kilo\parsec} beyond which the gradient plateaus at \SI{8.5}{\dex}.

\item We also searched for any azimuthal abundance variations across M101's disk. We did not find any strong variations along the spiral arms nor between arm/inter-arm regions. However, we did find that the west and southwest regions of M101 have significantly steeper radial gradients than elsewhere in the galaxy by $\sim$\SI{0.4}{\dex\perradius}. 

\item We discussed both models for M101's abundance gradient in the context of its interaction history and dynamics. simple exponential gradients are believed to be a feature of inside-out growth and independent of environment. Furthermore, previous spectroscopic analyses of M101 have supported a simple exponential gradient. However, given the strong NE/SW asymmetry of M101's disk, we encourage spectroscopic studies that sample \hii\ regions throughout the outer disk to confirm the azimuthal dependency of the radial gradient.

\item Broken exponential gradients are supported by recent studies of other galaxies, both individually and in large surveys \citep[e.g.][]{bresolin2009,sanchez2014}. While numerous physical processes have been proposed to explain the presence of an abundance break followed by flattening, we argue that the possibility of a radial break in M101's disk could be a result of the dynamical corotation barrier \citep{lepine2011,scarano2013} which results in the inner and outer portions of the disk evolving independently from one another. Furthermore, the flattening could be caused by a burst of star formation from M101's interaction with NGC~5474 enriching the outer disk. However, caution must be expressed here as well, since abundances estimated with the $R_{23}$ ratio have been known to produce false breaks \citep{pilyugin2003}. 

\end{enumerate}

\begin{acknowledgments}

This work was supported in part by a Towson Memorial Scholarship to R.G.\ and by grants to J.C.M.\ from the National Science Foundation (award 1108964) and the Mt.\ Cuba Astronomical Foundation. A.E.W\ acknowledges support from the STFC [ST/S00615X/1]. This publication makes use of data products from the Two Micron All Sky Survey, which is a joint project of the University of Massachusetts and the Infrared Processing and Analysis Center/California Institute of Technology, funded by the National Aeronautics and Space Administration and the National Science Foundation. R.G.\ would like to thank S.\ Linden for sharing their work on the dynamical modeling of the M101-NGC~5474 interaction. We also thank the anonymous referee for a detailed report that helped improve this paper.

\end{acknowledgments}

\facility{CWRU:Schmidt}

\software{\texttt{Astropy} \citep{astropy1,astropy2}, \texttt{Matplotlib} \citep{matplotlib}, \texttt{NumPy} \citep{numpy}, \acs{cigale} \citep{noll2009,boquien2019}, \texttt{SciPy} \citep{scipy}, \texttt{emcee} \citep{emcee}}, \texttt{piecewise-regression} \citep{piecewise}

\bibliography{strong_line_bibliography.bib}
\bibliographystyle{aasjournal.bst}

\end{document}